\documentclass[11pt,cite,a4paper]{article}

\pdfoutput=1

\usepackage{amsmath}
\usepackage{amssymb}
\usepackage{a4wide}
\usepackage{graphicx}

\numberwithin{equation}{section}

\begin{document}

\begin{titlepage}

\rightline{MIT-CTP 4304}
\vspace{2.5truecm}


\centerline{\Large \bf  Holographic evolution of the mutual information}

\vspace{1.5truecm}

 \centerline{Andrea Allais\footnote[1]{e-mail: a\_allais@mit.edu}  \,and\; Erik Tonni\footnote[2]{e-mail: tonni@mit.edu}}   
    
\vspace{1cm}

\centerline{{\it  Center for Theoretical Physics,}}
\vspace{.1cm}
\centerline{{\it  Massachusetts Institute of Technology,}}
\vspace{.1cm}
\centerline{{\it Cambridge, MA 02139, USA}}

\vspace{2.5truecm}

\centerline{\bf Abstract}
\vspace{.5truecm}

\noindent
We compute the time evolution of the mutual information in out of equilibrium quantum systems  whose gravity duals are Vaidya spacetimes in three and four dimensions, which describe the formation of a black hole through the collapse of null dust.
We find the holographic mutual information to be non monotonic in time and always monogamous in the ranges explored.
We also find that there is a region in the configuration space where it vanishes at all times.
We show that the null energy condition is a necessary condition for both the strong subadditivity of the  holographic entanglement entropy and the monogamy of the holographic mutual information.

\end{titlepage}

\section*{Introduction}

Entanglement entropy is a measure of the quantum entanglement in systems with many degrees of freedom which has been object of intense investigation in condensed matter, quantum information and quantum gravity.

Given a system whose total Hilbert space $H$ can be written as a direct product $H=H_A \otimes H_B$ and whose state is characterized by the density matrix $\rho$, one can define the reduced density matrix $\rho_A \equiv \textrm{Tr}_B \rho$ by tracing over the degrees of freedom of $B$. Then, the entanglement entropy $S_A$ is the corresponding Von Neumann entropy $S_A = - \textrm{Tr}_A (\rho_A \log \rho_A)$. The the situation mostly studied in the literature is when $A$ is given by a spatial region and $B$ is its complement. In this case the entanglement entropy is also called geometric entropy and $S_A$ behaves according to an {\it area law\,}: in $d>1$ spatial dimensions we have $S_A \propto \textrm{Area}(\partial A)/\epsilon^{d-1} + \dots$, where $\epsilon$ is the UV cutoff of the theory and the dots represent terms of higher order in $\epsilon$ \cite{Srednicki:1993im}. In two spacetime dimensions the divergence is logarithmic and more quantitative analysis has been performed for conformal field theories, where the symmetry provides powerful computational techniques. For these theories, when $A$ is given by an interval of length $\ell$ in an infinite line and the temperature is zero, we have the famous result $S_A = (c/3)\log (\ell/\epsilon)+c_1'$, where $c$ is the central charge and $c_1'$ is a constant \cite{Callan:1994py, Holzhey:1994we, Calabrese:2004eu} (see \cite{Calabrese:2009qy} for a recent review). The most useful method to get $S_A$ is the so called {\it replica trick}, which consists in computing the Renyi entropies 
$S^{(n)}_A \equiv (1-n)^{-1}\log \textrm{Tr} \rho_A^n$ for integer $n$. 
The entanglement entropy is then obtained as $S_A = - \partial_n \textrm{Tr} \rho_A^n |_{n=1}$.

For quantum field theories with a holographic dual, the prescription to compute $S_A$ in the boundary theory through a bulk computation has been conjectured in \cite{Ryu:2006bv,Ryu:2006ef} for static backgrounds and then generalized to stationary and time dependent geometries in \cite{Hubeny:2007xt} (see \cite{Nishioka:2009un} for a review). The proposal of  \cite{Ryu:2006bv,Ryu:2006ef} satisfies many properties and also important inequalities of the entanglement entropy (the simplest of them is the strong subadditivity) \cite{Hirata:2006jx, Headrick:2007km,Hayden:2011ag}. Nevertheless, a proof for this formula is not known \cite{Fursaev:2006ih,Headrick:2010zt}.

Interesting insights in the structure of entanglement can be obtained by considering two disjoint regions, namely $A=A_1 \cup A_2$ with $A_1 \cap A_2 = \emptyset$ \cite{Caraglio:2008pk, Furukawa:2008uk, Calabrese:2009ez, Calabrese:2010he}. In this case the most interesting quantity to study is the mutual information $I(A_1,A_2) \equiv S_{A_1} + S_{A_2} - S_{A_1 \cup A_2}$, which has the nice property that the leading divergence due to the area law cancels. The Renyi mutual information has been analytically computed for some simple two dimensional conformal field theories like the free compactified boson \cite{Calabrese:2009ez} and the Ising model \cite{Calabrese:2010he}, and it has been shown that it encodes all the data of the theory (all the conformal dimensions of the primaries and their correlation functions) in contrast with the entanglement entropy of a single interval which contains only the central charge. Unfortunately, the analytic continuation that allows to obtain $I(A_1, A_2)$ is not known in general but only in some limiting regimes. Detailed studies of $I(A_1, A_2)$ for spin chain models have also been done \cite{Alba:2009ek, Fagotti:2010yr, Fagotti:2010cc, Alba:2011fu}.

The holographic formula of \cite{Ryu:2006bv,Ryu:2006ef} for static backgrounds has been applied for disjoint regions \cite{Hubeny:2007re, Headrick:2010zt, Tonni:2010pv} and a qualitative disagreement   with respect to the expectations from the simple two dimensional conformal field theories mentioned above has been found. Indeed the holographic mutual information displays a continuos transition with discontinuous first derivative from zero to positive values as the two regions get closer. This could be explained through the fact that the holographic formula holds for large $c$ and it should be corrected in order to recover the results obtained for the models whose central charges are of the order of the unity. 
Nevertheless the holographic formula of \cite{Ryu:2006bv,Ryu:2006ef}  and its generalization for time dependent backgrounds \cite{Hubeny:2007xt} are believed to be correct for large $c$ and we will employ this prescription in our analysis.

The entanglement entropy is a very important quantity to study in order to understand the physics out of equilibrium and the processes of thermalization. In particular, one is interested in the unitary time evolution of the entanglement entropy when the system starts from a state which is not an eigenstate of the Hamiltonian of the system. This occurs for instance when a system is prepared in an eigenstate of the corresponding Hamiltonian and suddenly a tunable parameter of the Hamiltonian (e.g. the magnetic field) is changed ({\it global quench}). Then the system evolves unitarily accordingly with the new Hamiltonian starting from a state which is not one of its eigenstates. Another interesting situation is when the system is prepared in the ground state of two decoupled parts which are joined together at $t=0$ and then the whole system unitarily evolves through the translationally invariant Hamiltonian ({\it local quench}). These situations have been studied for two dimensional systems by employing conformal field theory  techniques and spin chains models \cite{Calabrese:2005in, DeChiara:2005wb, Calabrese:2006rx, Calabrese:2007rg, Calabrese:2007xx}.

Thermalization processes have been also widely considered from the holographic point of view (see \cite{Hubeny:2010ry} for a review). For instance, the holographic counterpart of the unitary evolution of a system towards a stationary state is a gravitational collapse whose final state is a black hole. In this process both the initial and the final states are thermal and the holographic entanglement entropy can be computed through the prescription of  \cite{Hubeny:2007xt}.\\
The Vaidya metrics (see e.g. \cite{Stephani2003}) are simple backgrounds realizing this setup.  
They are solutions of the Einstein equations with negative cosmological constant and a non trivial energy momentum tensor containing a mass function constrained by the null energy condition which
describe the formation of a black hole through the collapse of a shell of null dust.
The null geodesics in these geometries have been studied in \cite{Hubeny:2006yu} but here we are interested into the spacelike ones, which occur in the computation of the holographic entanglement entropy \cite{Hubeny:2007xt, AbajoArrastia:2010yt, Albash:2010mv}.
The Vaidya metrics are simplifications of more general models considered e.g. in \cite{Lin:2008rw} (tensionful shell) and \cite{Bhattacharyya:2009uu}. Other holographic thermalization setup have been also studied \cite{Das:2010yw}. The study of the holographic entanglement entropy in Vaidya backgrounds is usually numerical but recently  an analytic computation has been done for the limiting regime of thin shell in three spacetime dimensions \cite{Balasubramanian:2010ce, Balasubramanian:2011ur}. We will largely employ this result in our analysis because it allows to explore a larger range of parameters. The holographic analysis of the two point functions in these dynamical geometries has been carried out in \cite{Balasubramanian:2010ce, Balasubramanian:2011ur, Aparicio:2011zy}.\\

In this paper we study the holographic entanglement entropy in the three and four dimensional Vaidya backgrounds.  
In section \ref{section vaidya metric} we introduce the metrics and the corresponding holographic entanglement entropy for a single region in the boundary theory. In section \ref{section HMI} we study the holographic mutual information and its transition curves in the configuration space. In section \ref{section SSA and NEC} we explore the relation between the null energy condition for the Vaidya metrics and the strong subadditivity for the holographic entanglement entropy. In section \ref{section I3} we extend the analysis performed in the previous sections to the holographic tripartite information in order to verify the monogamy of the holographic mutual information in Vaidya spacetimes and study this property is influenced by the null energy condition.

\noindent {\bf Note added.} While we were completing the writing of this paper, \cite{Balasubramanian:2011LAST} appeared and it has a substantial overlap with our results.

\section{Holographic entanglement entropy for Vaidya geometries}
\label{section vaidya metric}

In this section we introduce the Vaidya metrics in $d+1$ dimensions (subsection \ref{subsection vaidya}) and we describe some known results about the holographic entanglement entropy in these backgrounds (subsection \ref{subsection HEE vaidya}). In subsection \ref{section 3d thin} we focus on the three dimensional case because it is the simplest to study and some analytical results have recently been found \cite{Balasubramanian:2010ce, Balasubramanian:2011ur} which will be widely employed in the remaining part of the paper.

\subsection{Vaidya metrics}
\label{subsection vaidya}

The $d+1$ dimensional Vaidya metrics in the Poincar\'e coordinates read
\begin{equation}
\label{metric vadya d dim}
ds^2
= \frac{l^2}{z^2}
\Big[ -\big(1-m(v) z^{d}\big)dv^2-2 dz  dv + d\vec{x}^2\,\Big]
\end{equation}
(we set $8\pi G_N^{(d+1)} =1$) where $\vec{x}=\{x_1, \dots ,x_{d-1}\}$ are the spatial boundary coordinates. The Ricci scalar of (\ref{metric vadya d dim}) is $R= -(d+1)d/l^2$.
The metric (\ref{metric vadya d dim}) is a solution of the Einstein equations in presence of matter
 \begin{equation}
 \label{eom EH}
G_{\mu\nu} + \Lambda g_{\mu\nu} = T_{\mu\nu}
\hspace{2cm}
\Lambda = - \frac{d(d+1)}{2l^2}
\end{equation}
where the energy momentum tensor has only one non vanishing component
 \begin{equation}
 \label{Tmunu}
T_{vv} = \frac{d-1}{2}\,z^{d-1} \partial_v m(v)\;.
\end{equation}
The metric (\ref{metric vadya d dim}) describes the formation of a black hole through the collapse of null dust, which is characterized by the $T_{\mu\nu}$ just introduced.
It is important to observe that the translational invariance along the directions contained in $\vec{x}$ is preserved at any time. This is a key feature of the setup characterizing global quenches in the boundary theory.\\
In processes of gravitational collapse, it is not yet understood how to characterize the formation of a black hole through local time evolution. To this purpose, some generalizations of the concept of horizon have been proposed, and, in our backgrounds we can distinguish between two horizons,
the event and the apparent horizon \cite{HawkingEllis, Wald, Booth:2005qc}. 
The apparent horizon is the boundary  of the trapped surfaces associated to a given foliation. For the metrics (\ref{metric vadya d dim}) a foliation which preserves the translation invariance in the directions of $\vec{x}$ is given by $v= \textrm{const}$ and $z= \textrm{const}$.
The location of the apparent horizon of (\ref{metric vadya d dim}) reads \cite{Hubeny:2007xt, AbajoArrastia:2010yt}
 \begin{equation}
 \label{app hor}
z_a = \frac{1}{m(v)^{1/d}}\;.
\end{equation}
Instead, the event horizon is given by 
 \begin{equation}
  \label{event hor}
\frac{dz_e}{dv} \,=\,-\frac{1-m(v) z_e^{d}}{2}\;.
\end{equation}
 When the mass profile $m(v)$ is constant $m(v)=M$, the metric (\ref{metric vadya d dim}) describes the geometry of the Schwarzschild black hole with planar horizon. This can be clearly seen through the following change of coordinates
 \begin{equation}
v\,=\,t +p(z)
\hspace{2cm}
p'(z) = -\frac{1}{1-M z^{d}}
\end{equation}
which allows to write (\ref{metric vadya d dim}) as 
\begin{equation}
\label{Schwarzschild BH}
ds^2
= \frac{l^2}{z^2}
\left[ -\big(1-M z^{d}\big)dt^2+ \frac{dz^2}{1-M z^{d}} + d\vec{x}^2\,\right]
\end{equation}
i.e. the usual form for a Schwarzschild black hole of mass $M$ in the Poincar\'e coordinates. For $l=1$
the Hawking temperature of this black hole is given by $T_H=  d M^{1/d}/(4\pi)$.
It is also straightforward to check that for $m(v)=0$ identically the metric (\ref{metric vadya d dim}) describes $AdS_{d+1}$ in the Poincar\'e coordinates
\begin{equation}
\label{ads metric}
ds^2 \,=\,\frac{l^2}{z^2}\big( -dt^2+ dz^2 + d\vec{x}^2\big)
\hspace{2cm}
t\,=\,v+z\;.
\end{equation}
This tells us that the Vaidya metric (\ref{metric vadya d dim}) is asymptotically $AdS_{d+1}$. 
\begin{figure}[h]
\begin{center}
\includegraphics[width=5.2in]{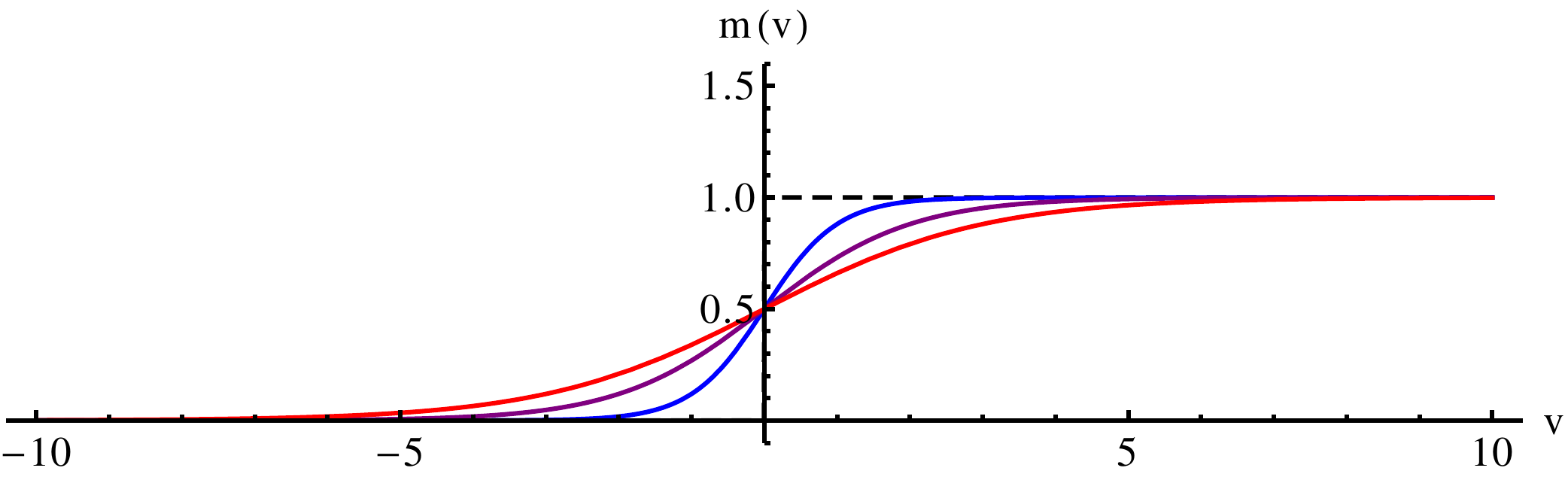}
\end{center}
\vspace{-.7cm}
\caption{The function (\ref{mass pos kink}) for different values of the thickness $a_v$ and $M=1$. The dashed curve is a step function and corresponds to the the limiting regime of thin shell $a_v \rightarrow 0$.
\label{plot mass upkink}}
\end{figure}

\noindent
The formation of the Schwarzschild black hole (\ref{Schwarzschild BH}) from $AdS_{d+1}$ is described by considering a function $m(v)$ which interpolates between $0$ and the finite value $M>0$ in an strictly increasing way. The profile for $m(v)$ is usually chosen to be
\begin{equation}
\label{mass pos kink}
m(v) = M\,\frac{1+\tanh(v/a_v)}{2}\;.
\end{equation}
Given this $m(v)$, the metric (\ref{metric vadya d dim}) describes a background which evolves from pure planar $AdS_{d+1}$ at early times to Schwarzchild black brane with mass $M$ at late times because of the infalling shell of null dust.
The parameter $a_v$ determines the transition between these two regimes since it parameterizes the thickness of the shell which falls along $v=0$. The mass function (\ref{mass pos kink}) is shown in the figure \ref{plot mass upkink}.
In the limiting case of $a_v \rightarrow 0$, the mass profile  $m(v)$ becomes a step function $M \theta(v)$ and the infalling shell describes a shock wave. This limit is very interesting because it captures the essential physics of the problem and one can hope to find analytic solutions for the quantities considered. For the holographic entanglement entropy in three spacetime dimensions this has been done in \cite{Balasubramanian:2010ce, Balasubramanian:2011ur} and we will largely employ this result.\\
An important inequality to impose on the energy momentum tensor $T_{\mu\nu}$ in (\ref{eom EH}) to guarantee the positivity of the energy density is the {\it null energy condition}, namely $T_{\mu\nu} N^\mu N^\nu \geqslant 0$ for any null vector $N^\mu$ \cite{HawkingEllis, Wald}. 
This inequality has been employed to study the $c$ theorems from the holographic point of view \cite{Freedman:1999gp, Myers:2010tj} and their generalizations in presence of boundaries \cite{Affleck:1991tk, Karch:2000ct, Takayanagi:2011zk, Fujita:2011fp}.
For the energy momentum tensor (\ref{Tmunu}), imposing the null energy condition means 
\begin{equation}
\label{pos mprime}
\partial_v m(v) \geqslant 0
\end{equation}
which is clearly satisfied by the profile (\ref{mass pos kink}). In the sections \ref{section SSA and NEC} and \ref{section I3} we consider mass profiles violating this condition and the effect of this violation on the holographic entanglement entropy.

\subsection{Holographic entanglement entropy}
\label{subsection HEE vaidya}

For static spacetimes like $AdS_{d+1}$ (\ref{ads metric}) and the Schwarzschild black hole (\ref{Schwarzschild BH}), the prescription to obtain the entanglement entropy $S_A = - \textrm{Tr}(\rho_A \log \rho_A)$ in the boundary theory through a holographic computation in the bulk has been proposed in \cite{Ryu:2006bv, Ryu:2006ef}. It reads
\begin{equation}
\label{RT formula}
S_A \,=\,\frac{\textrm{Area}(\gamma_A)}{4 G_N^{(d+1)}}
\end{equation}
where $G_N^{(d+1)}$ is the Newton constant for the $d+1$ dimensional bulk and  $\gamma_A$ is defined as the {\it minimal} surface among the spatial ones which extend into the bulk and share the boundary with $A$, i.e. $\partial \gamma_A = \partial A$ ($\gamma_A$ is homologous to $A$). Thus, $\gamma_A$ is a codimension two surface living in the $t = \textrm{const}$ slice and has minimal area. 
Since $\gamma_A$ lives in an asymptotically $AdS_{d+1}$ space and it reaches its boundary, placed at $z=0$ in the Poincar\'e coordinates, $\textrm{Area}(\gamma_A)$ is infinite and therefore it must be regularized by introducing a small cutoff $\epsilon>0$ in the holographic direction, namely the restriction $z > \epsilon$ is imposed during the bulk computation. The series expansion of $S_A$ in $\epsilon$ depends on $d$ and one the first checks of (\ref{RT formula}) was that this expansion
reproduces in the holographic way the leading UV divergence of the entanglement entropy which was computed in field theory through other methods. 
In particular, for $d>2$, the leading divergence of $S_A$ is proportional through a non universal coefficient to $\textrm{Area}(\partial A)/\epsilon^{d-2}$ \cite{Srednicki:1993im} (this is the so called {\it area law}). For $d=2$, when the boundary CFT is two dimensional and the spatial region $A$ is a segment of length $\ell$ (thus $\partial A$ is made by two points), the entanglement entropy $S_A$ diverges logarithmically with a universal coefficient given by the central charge of the theory \cite{Callan:1994py, Holzhey:1994we, Calabrese:2004eu}.
Besides these fundamental checks, it has been shown that the holographic proposal (\ref{RT formula}) satisfies the strong subadditivity condition \cite{Headrick:2007km} and other more complicated inequalities characterizing the entanglement entropy \cite{Hayden:2011ag}, which will be discussed in sections \ref{section SSA and NEC} and \ref{section I3}.

The proposal (\ref{RT formula})  for static backgrounds has been generalized
to time dependent geometries  in \cite{Hubeny:2007xt}. In these cases $S_A$ is still given by (\ref{RT formula}) but with $\gamma_A$ obtained as an {\it extremal} surface, namely as the saddle point of the proper area functional. This proposal is covariantly well defined and reduces to the previous one when the spacetime is static.
Let us explain it in details for the Vaidya $d+1$ dimensional spacetimes (\ref{metric vadya d dim}).\\
We consider as $d-1$ dimensional spatial region $A$ in the boundary theory a rectangular strip parameterized by $x_1 \in (- \ell/2,\ell/2)$ and $x_2, \dots x_{d-1} \in (0, \ell_\perp)$ at fixed value of the boundary time coordinate $t$.
This choice is less symmetric than the case of $A$ given by a circular shape but a crucial simplification occurs in this case.\\
According to the proposal of  \cite{Hubeny:2007xt}, the holographic entanglement for this spatial region $A$ is given by the area of the extremal surface $\gamma_A$ whose profile is most conveniently specified by $v\equiv v(x_1)$ and $z\equiv z(x_1)$ (we are assuming that $\gamma_A$ is translationally invariant in the other boundary coordinates $x_2, \dots x_{d-1}$ parameterizing $A$) with the following boundary conditions
 \begin{equation}
 \label{bc conds}
v(-\ell/2) = v(\ell/2)  = t
\hspace{2cm}
z(-\ell/2) = z(\ell/2)  = 0\;.
\end{equation}
These boundary conditions impose that the boundary of $\gamma_A$ coincides with the boundary of $A$ along the boundary temporal evolution.
Since $x_1$ is the relevant independent variable, we will denote it by $x$ in the following. The area of the such {\it spacelike} surface (thus the determinant of the induced metric under the square root must be taken with the positive sign) is given by the following functional
 \begin{equation}
 \label{area d dim}
\textrm{Area}(\gamma_A) \,\equiv\, \mathcal{A}_d 
\,\equiv\, l^{d-1}\,2\ell_\perp^{d-2}
\int_{0}^{\frac{\ell}{2}}
\frac{1}{z^{d-1}} \sqrt{1-\big[1-m(v)z^d\big](v')^2-2v' z'}\,dx
\end{equation}
where $' \equiv d/dx$ and we have chosen the origin along the $x$ direction in order to employ the fact that the functions $v(x)$ and $z(x)$ are even. This consideration about parity determines the factor 2 and the integration extrema in (\ref{area d dim}).\\
The area functional (\ref{area d dim}) is the one we have to extremize in order to get the the codimension two surface $\gamma_A$ allowing to compute the holographic entanglement entropy for the time dependent Vaidya spaces through (\ref{RT formula}) \cite{Hubeny:2007xt}.
In other words, $\gamma_A$ is a solution of the two equations of motion of (\ref{area d dim}).
Since the integrand in (\ref{area d dim}) does not contain $x$ explicitly, we have the following conservation equation
 \begin{equation}
 \label{cons eq}
\left(\frac{z_\ast}{z}\right)^{2(d-1)} =\,
1-\big[\,1-m(v)z^d\,\big](v')^2-2v' z'
\end{equation}
where $z_\ast = z(0)$ is the maximum value of $z(x)$, which is characterized by $z'=v'=0$. 
The occurrence of this conservation law is the simplification characterizing for the rectangular shape of $A$ with respect to the circular one mentioned above.
The two equations of motion obtained by minimizing the functional (\ref{area d dim}) read
\begin{eqnarray}
\label{eom d dim v}
\big[1-m(v)z^d\big]v''+z''  -\frac{\partial_v m(v)}{2}\,z^d (v')^2 -d\, m(v) z^{d-1} z' v'
&=& 0
\\
\label{eom d dim z}
\rule{0pt}{.6cm}
z\, v''-\frac{d-2}{2}\,m(v) z^d (v')^2+(d-1)\big[\,(v')^2+2v' z'-1\,\big]
&=& 0\;.
\end{eqnarray}
By taking the derivative w.r.t. $x$ of the conservation equation (\ref{cons eq}) and using one of these two equations of motion, one obtains the other one. Thus, it is sufficient to consider only (\ref{cons eq}) and e.g. (\ref{eom d dim z}) to find $v(x)$ and $z(x)$.\footnotetext{${^\dagger}$In all the plots of this paper the numerical values of the parameter characterizing the different curves are evenly spaced.}
\begin{figure}[th]
\vspace{.5cm}
\begin{tabular}{ccc}
\hspace{-.8cm}
\includegraphics[width=3.2in]{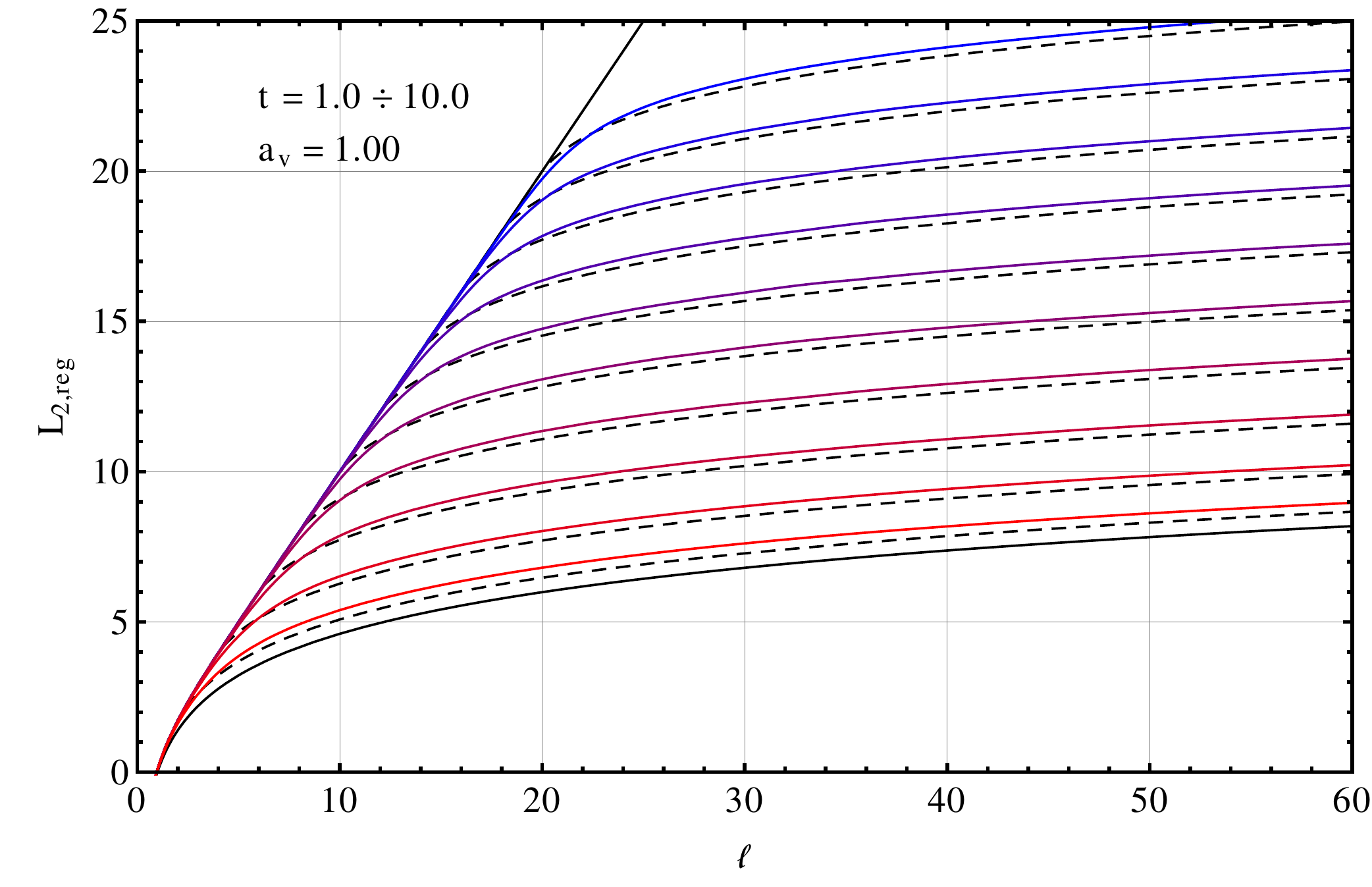}
& \hspace{-.7cm} & 
\includegraphics[width=3.2in]{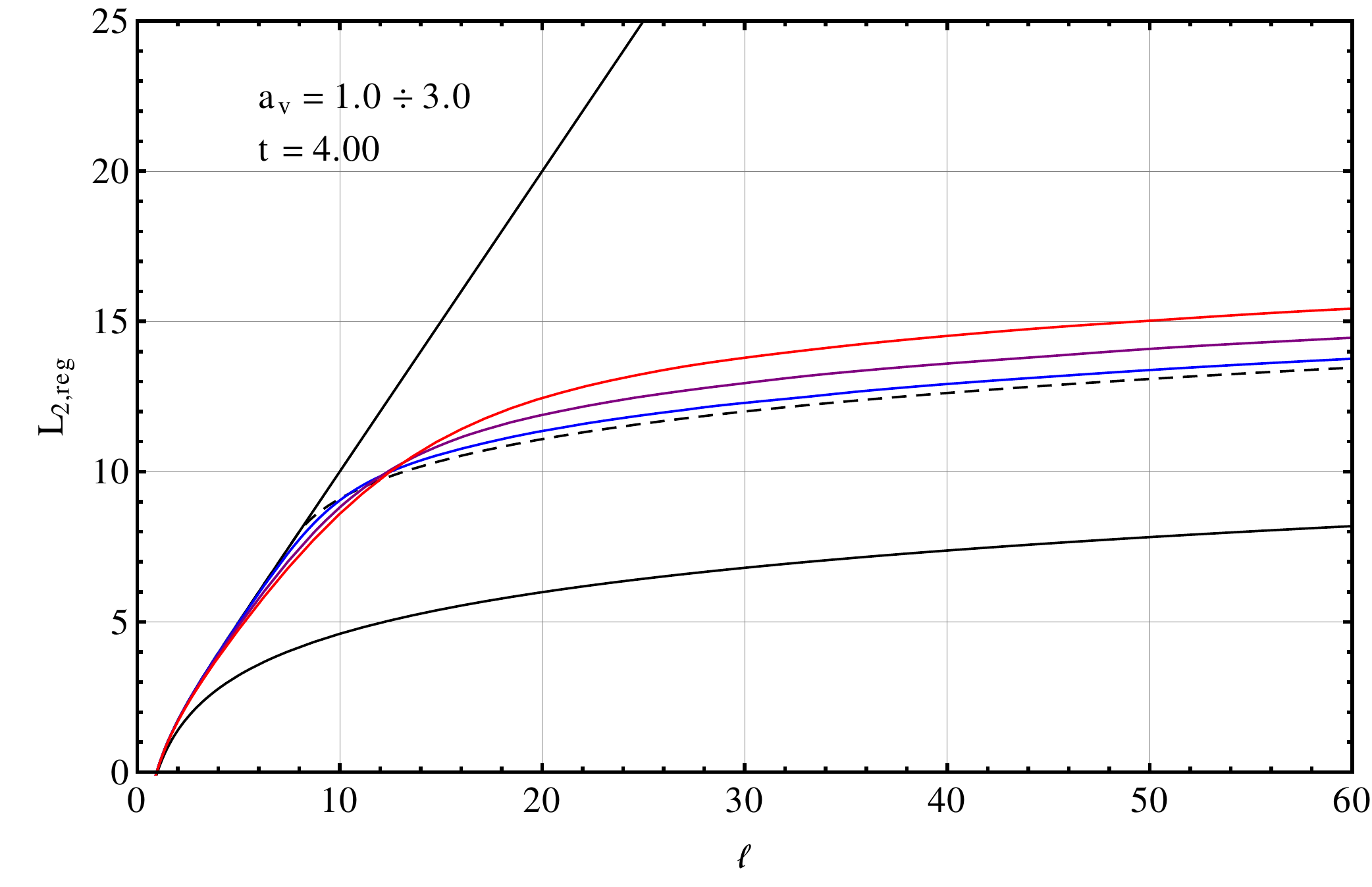}
\end{tabular}
\vspace{-.6cm}
\caption{Finite term of the holographic entanglement entropy for the Vaidya metric in three dimensions (see (\ref{RT formula}) and (\ref{Ad reg def}) up to a factor) with $m(v)$ given by (\ref{mass pos kink}) \cite{Hubeny:2007xt, AbajoArrastia:2010yt}. On the left, $L_{2,\textrm{reg}}(\ell,t)$ for different values${^\dagger}$ of  boundary time $t$, increasing from the red curve to the blue one, at a fixed value of $a_v$. On the right, $L_{2,\textrm{reg}}(\ell,a_v)$ at a fixed value of $t$ and different values of the thickness $a_v$, which increases going from the blue curve to the red one. 
In both the plots, the black curves correspond to the limiting regimes of $AdS_3$ (bottom curve, from (\ref{Lreg ads3 btz})) and of the BTZ blak hole (top curve, from (\ref{Lreg ads3 btz})), while the dashed curves represent the corresponding curves for the thin shell limit $a_v=0$.
\label{plot Lregposkink}}
\end{figure}

\begin{figure}[h]
\begin{center}
\vspace{.5cm}
\includegraphics[width=4.2in]{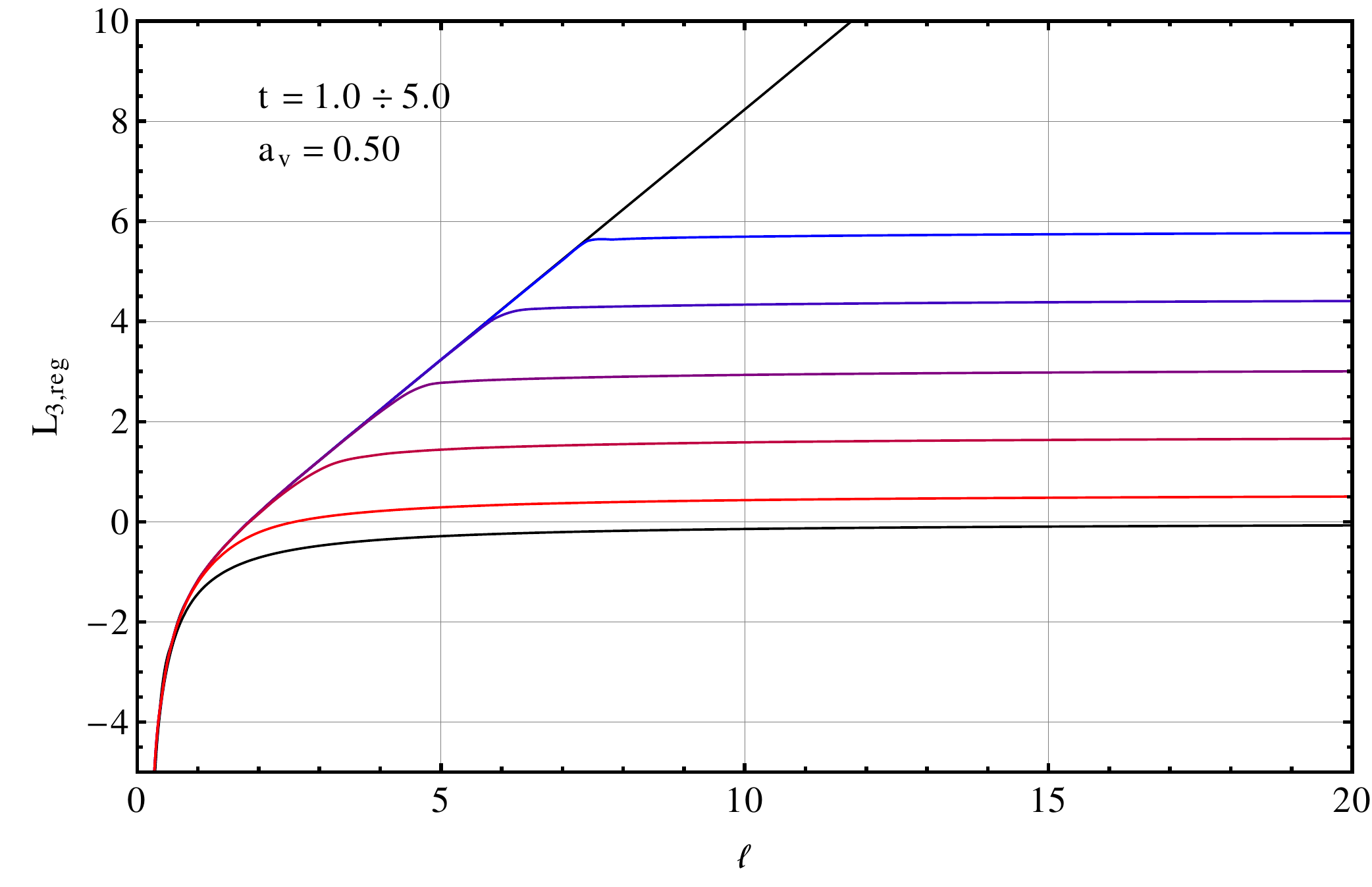}
\end{center}
\vspace{-.7cm}
\caption{Finite term of the holographic entanglement entropy for the Vaidya metric in four 
dimensions (proportional to $L_{3,\textrm{reg}}(\ell,t)$) with $m(v)$ given by (\ref{mass pos kink}), fixed $a_v$ and different boundary times $t$, increasing from the red curve to the blue one. The black curves correspond to the limiting regimes of $AdS_4$ (bottom curve from (\ref{Lreg ads})) and of the Schwarzschild black hole in four dimensions (top curve).
\label{plot Lregposkink d=3}}
\end{figure}

\noindent 
Given a solution $(v(x),z(x))$ of the equations of motion, its area can be computed by evaluating the integral (\ref{area d dim}) on it.  Using (\ref{cons eq}), this area can be written as follows
 \begin{equation}
\mathcal{A}_d = l^{d-1}\,2\ell_\perp^{d-2}
\int_{0}^{\frac{\ell}{2}}
\frac{z_\ast^{d-1}}{z^{2(d-1)}}\,dx\;.
\end{equation}
As discussed above, this integral is divergent because the spacetime we are dealing with, being asymptotically $AdS_{d+1}$, is non compact and the spatial surface $\gamma_A$ we are considering reaches its boundary (see the boundary conditions (\ref{bc conds})).\\
The divergence of $\mathcal{A}_d$ can be obtained by studying the same problem in $AdS_{d+1}$ in the standard way \cite{Ryu:2006ef}. Subtracting this divergence we obtain the finite term of the area which is the main quantity we are interested in. For $d>2$ we have
 \begin{equation}
 \label{Ad reg def}
\mathcal{A}_{d,\textrm{reg}} \,\equiv\, 2l^{d-1}\,\ell_\perp^{d-2}
  \lim_{\eta \,\rightarrow \,0^+} 
\left(\,
\int_{0}^{\frac{\ell}{2}-\eta}
\frac{z_\ast^{d-1}}{z^{2(d-1)}}\,dx
-\frac{1}{(d-2)\, \epsilon^{d-2}}
\right)
\,\equiv\, l^{d-1}\,\ell_\perp^{d-2} L_{d,\textrm{reg}}
\end{equation}
where $\epsilon \equiv z(\ell/2-\eta)$ is the UV cutoff in the boundary theory.
Notice that in this case we need to subtract just one diverging term to regularize $\mathcal{A}_{d}$. This is a feature of the strip; indeed when the region $A$ is a circle there are more terms to subtract to make the area finite \cite{Ryu:2006ef}.

In order to find the solution of (\ref{cons eq}) and (\ref{eom d dim z}) satisfying the boundary conditions  (\ref{bc conds}), first we exploit the reflection symmetry about $x = 0$ and solve the Cauchy problem whose initial conditions are given by
\begin{equation}
 z(0) = z_*
\hspace{2cm}
 v(0) = v_*\;.
\end{equation}
Then, we shoot in the variables $z_*$, $v_*$ to impose (\ref{bc conds})\footnote{This approach is different from the one used in \cite{AbajoArrastia:2010yt}, where shooting takes place from the boundary. We favored shooting from $x = 0$ because it is a regular point of the solution $(v(x), z(x))$.}. It turns out that the points at which the solution reaches the boundary become increasingly sensitive to the initial conditions and to the accuracy of the integration as either $d$ or $\ell$ increase. As a consequence, it becomes more and more difficult to impose (\ref{bc conds}). This technical difficulty limits  the range of parameters we are able to explore for $d > 2$. Once the solution is found, the implementation of the numerical integration and limit in (\ref{Ad reg def}) is quite straightforward.

In the figures \ref{plot Lregposkink} and \ref{plot Lregposkink d=3} we show respectively  $L_{2,\textrm{reg}}$ (three dimensional Vaidya) and $L_{3,\textrm{reg}}$ (four dimensional Vaidya)  as functions of $\ell$ and for different values of the two other important parameters involved in our problem: the boundary time $t$ and the thickness of the shell $a_v$. The three dimensional case deserves a separated discussion (see subsection \ref{section 3d thin}) because it is the simplest situation and therefore some  analytic results can be found \cite{Balasubramanian:2010ce, Balasubramanian:2011ur}. Notice that the main qualitative features of the plots in the figures \ref{plot Lregposkink} and \ref{plot Lregposkink d=3} are independent of the number of dimensions.
The black curves represent the limiting regimes, which are $AdS_{d+1}$ in the early times (bottom curve) and the $d+1$ dimensional Schwarzschild black hole at late times (top curve). For generic $d>2$ the result for $AdS_{d+1}$ is known \cite{Ryu:2006ef} 
 \begin{equation}
  \label{Lreg ads}
L_{d,\textrm{reg}} \,\Big|_{AdS_{d+1}} =
\,-\frac{(2\sqrt{\pi})^{d-1}}{d-2} \left[\,\frac{\Gamma(\tfrac{d}{2(d-1)})}{\Gamma(\tfrac{1}{2(d-1)})}\,\right]^{d-1}
\frac{1}{\ell^{d-2}}\;.
\end{equation}
Unfortunately $L_{d,\textrm{reg}}$ for the $d+1$ dimensional Schwarzschild black hole is not known. Very few analytic results are available for minimal surfaces in four and higher dimensional black holes but the curves for $L_{d,\textrm{reg}}$ have been studied \cite{Barbon:2008sr, Tonni:2010pv}.\\
At any intermediate, finite and fixed boundary time $t$ during the black hole formation, we can observe from the figures \ref{plot Lregposkink} and \ref{plot Lregposkink d=3} that $L_{d,\textrm{reg}}(\ell,t)$ goes over the  Schwarzschild black hole curve for small $\ell$ and at some point (which depends on $t$) it leaves from it to adopt a $AdS_{d+1}$ like behavior shifted vertically.
Indeed, at any finite time the shell is fixed in some region of the bulk and we have a $AdS_{d+1}$ geometry inside the shell and a  Schwarzschild black hole outside it, with the thickness of the transient region is parameterized by $a_v$. For small values of $\ell$, the extremal surface stays completely outside the shell and far from it feeling therefore only the Schwarzschild black hole geometry.
As $\ell$ increases, the extremal surface begins to enter into the interior part of the shell and therefore to feel the $AdS_{d+1}$ geometry inside. This makes $L_{d,\textrm{reg}}$ deviate from the Schwarzschild behavior. 
When $\ell$ is very large, a big part of the extremal surface is inside the shell and therefore its length is determined by $AdS_{d+1}$, explaining the asymptotic behavior of the curves in the figures \ref{plot Lregposkink} and \ref{plot Lregposkink d=3} for large $\ell$. The vertical shift in this regime is due to the part of the surface which is close to the boundary: being outside the shell, it feels the Schwarzschild black hole geometry and it provides a larger contribution to $L_{d,\textrm{reg}}$ than the $AdS_{d+1}$ geometry. 
From the plot on the right in the figure \ref{plot Lregposkink}, we can observe that as the thickness $a_v$ decreases, $L_{d,\textrm{reg}}$ reproduces the Schwarzschild black hole result for a larger range of $\ell$ and also that the vertical shift for large values of $\ell$ w.r.t. the $AdS_{d+1}$ curve decreases.\\
At this point we find it useful to have a look to the shape of the extremal surfaces, which is shown in the figure \ref{plot geodesics} for the case of two disjoint regions in the boundary and will be studied in the section \ref{section HMI}. The extremal surfaces go backward in the bulk time direction $v$ and around $v=0$ they penetrate into the shell probing the $AdS_{d+1}$ geometry inside 
\cite{AbajoArrastia:2010yt}.

\subsection{Three dimensional case and thin shell limit}
\label{section 3d thin}

The three dimensional Vaidya geometry ($d=2$) is the simplest situation and consequently the best one to study in order to get a physical intuition and more analytic results \cite{Hubeny:2007xt, AbajoArrastia:2010yt, Balasubramanian:2010ce, Balasubramanian:2011ur} that could be helpful for the higher dimensional case.
Moreover, the CFT on the boundary theory is two dimensional and the powerful methods developed for these class of models lead to important results \cite{Callan:1994py, Holzhey:1994we, Calabrese:2004eu} which are very helpful to test the holographic techniques.

\begin{figure}[th]
\vspace{.4cm}
\begin{tabular}{ccc}
\vspace{.4cm}
\hspace{-.6cm}
\includegraphics[width=3.2in]{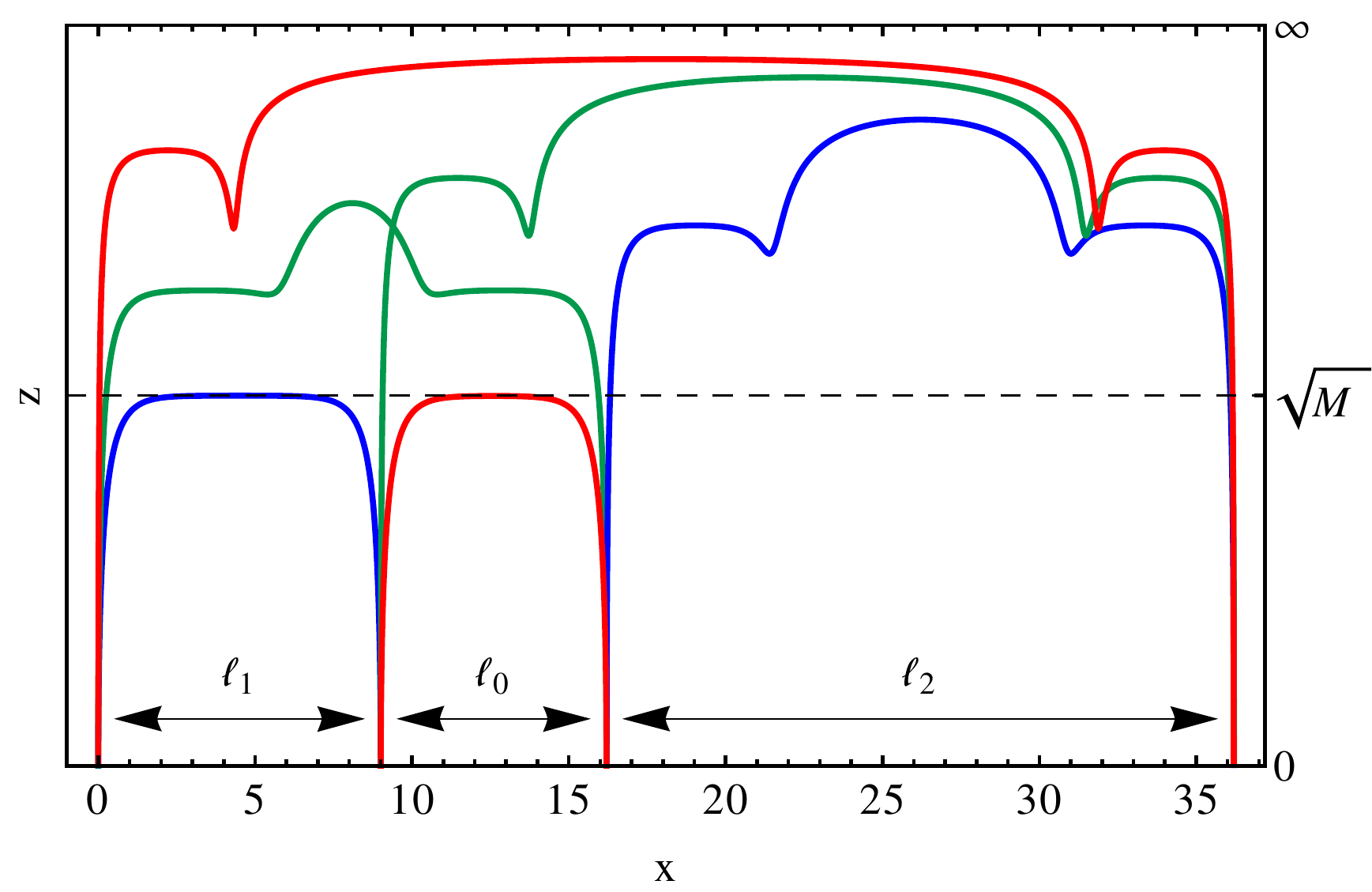}
& \hspace{-.7cm} & 
\includegraphics[width=3.2in]{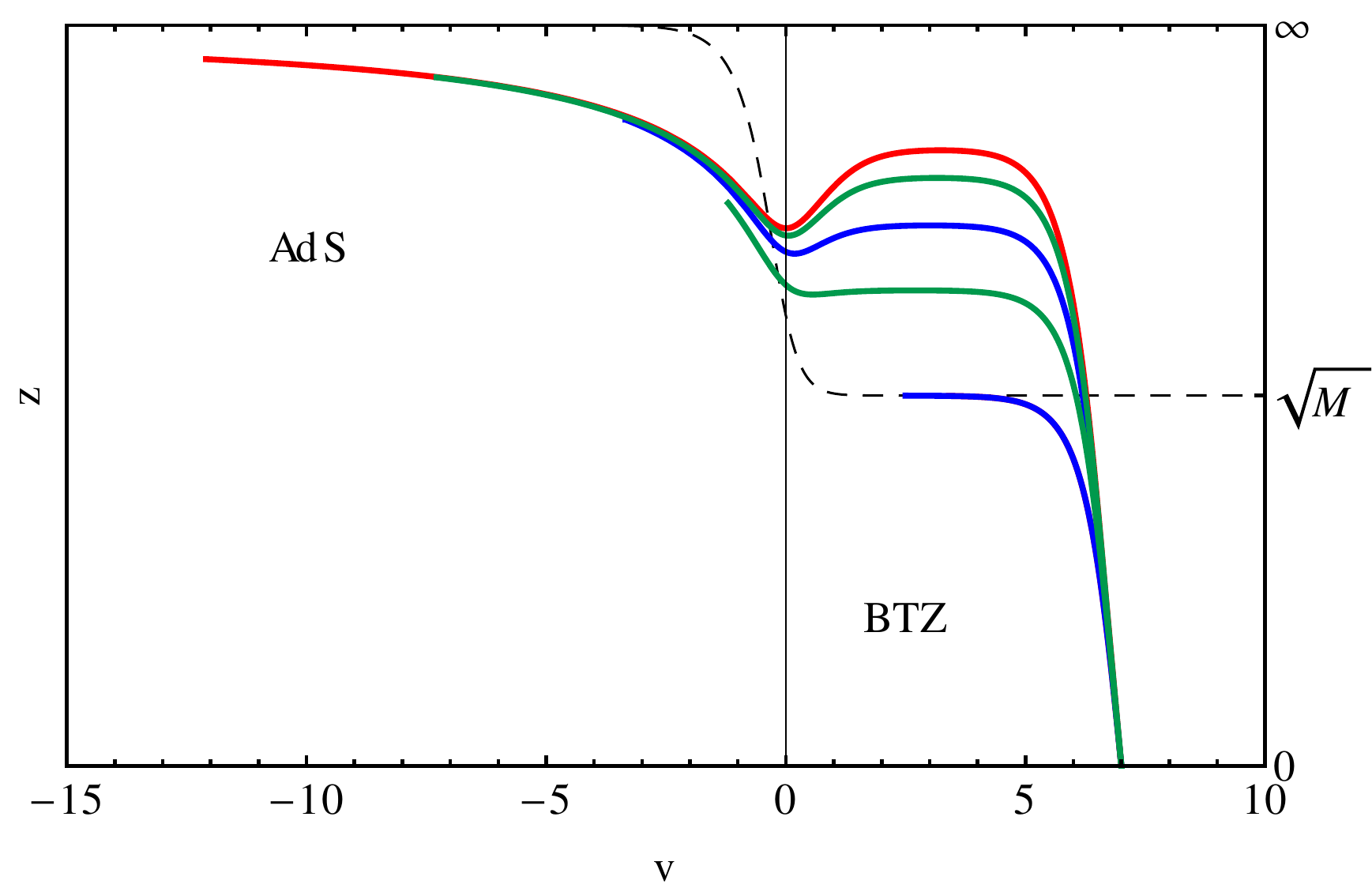}\\
\vspace{.2cm}
\hspace{-.6cm}
\includegraphics[width=3.2in]{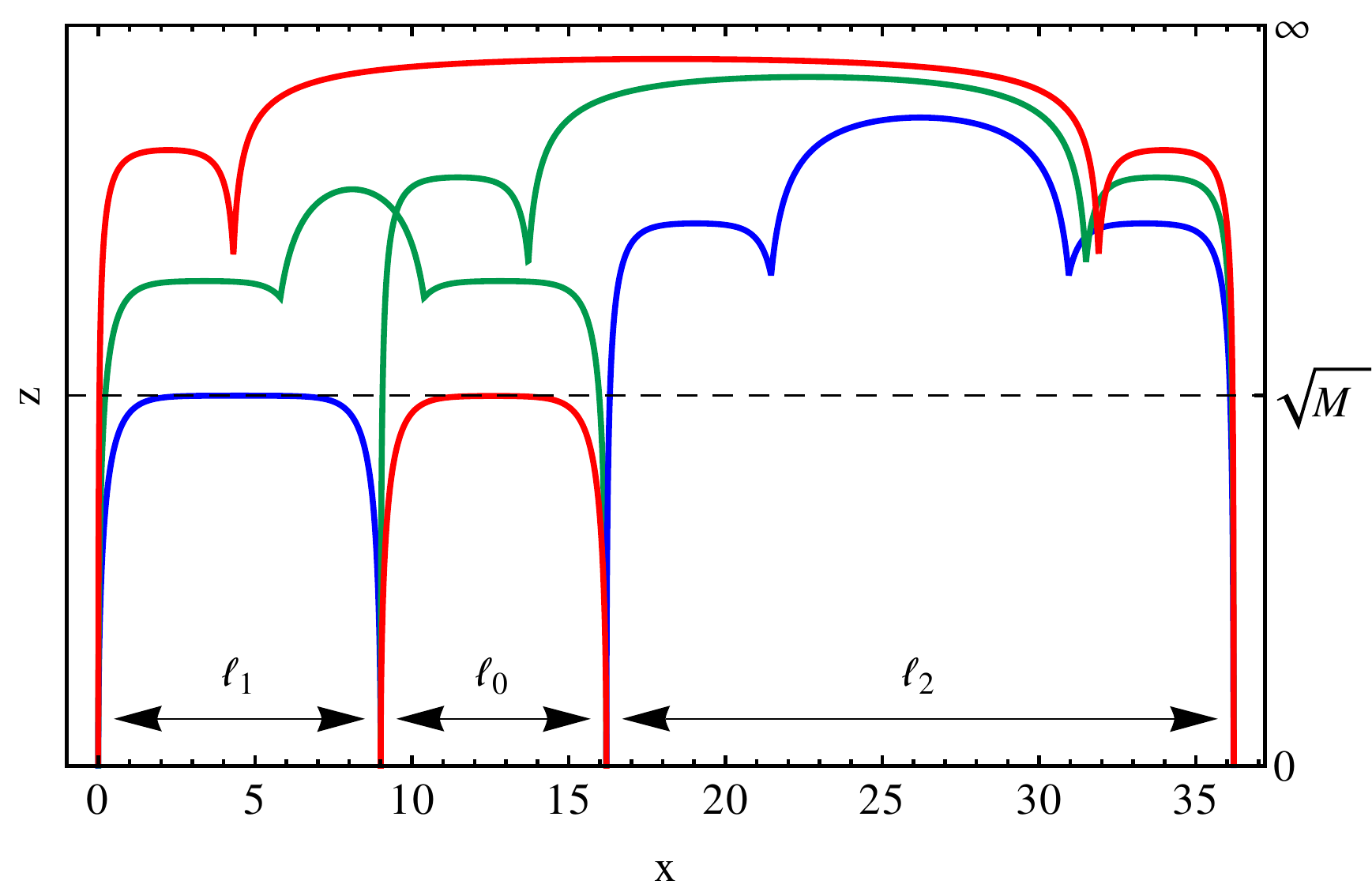}
& \hspace{-.7cm} & 
\includegraphics[width=3.2in]{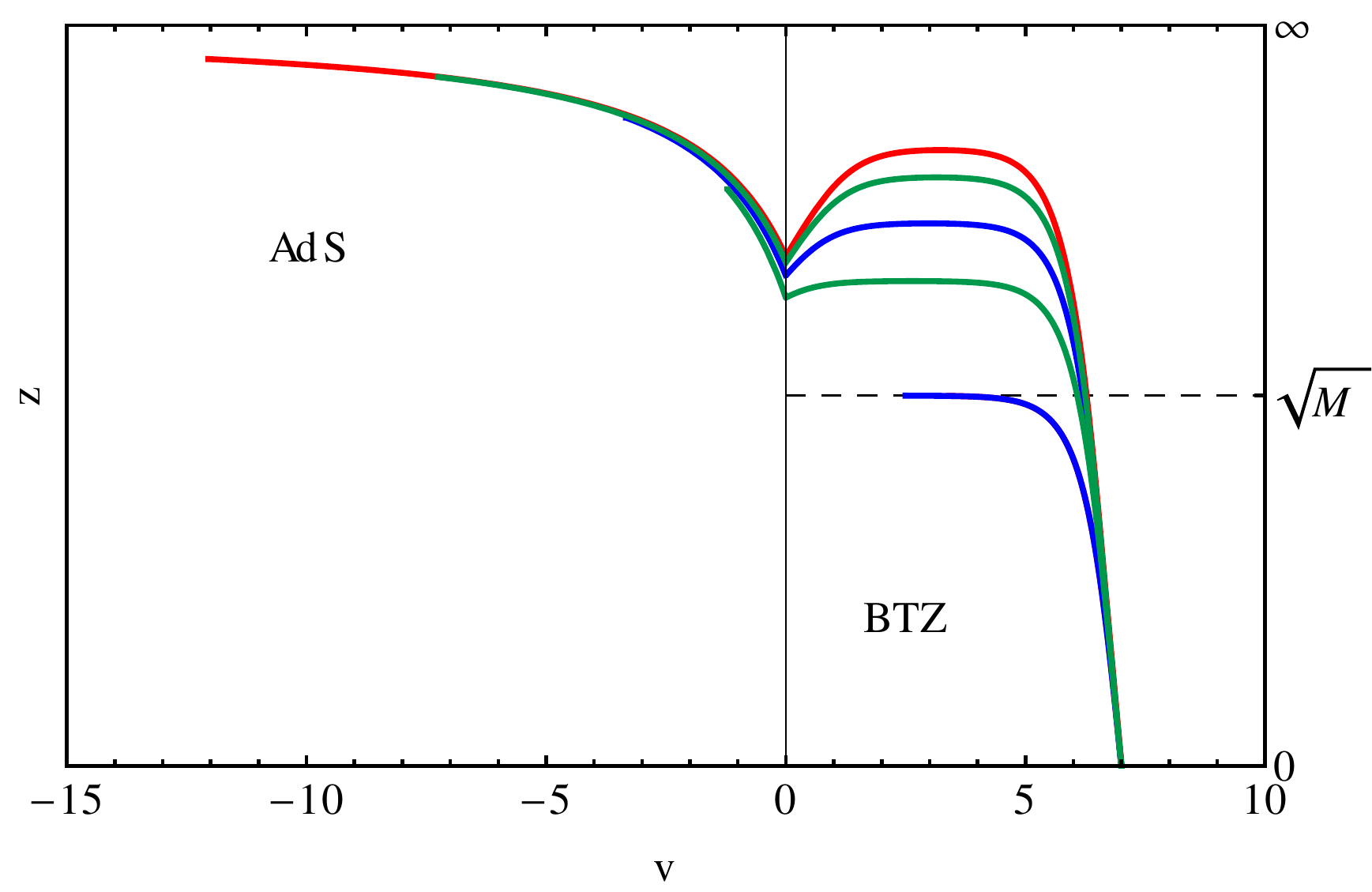}
\end{tabular}
\caption{Geodesics configuration for the holographic mutual information at the transition point for three dimensional Vaidya geometry ($d=2$) with $M=1$ in (\ref{mass pos kink}). The total length of the geodesics for the connected configuration (red) and the disconnected one (blue) is the same.
In the upper plots $a_v = 0.5$ while in the bottom ones $a_v \rightarrow 0$ (thin shell limit). The  boundary time is $t = 7$ (see the intersection of the curves with the horizontal axis in the plots on the right). The $z$ axis has been compactified using the $\arctan$ function. The green geodesics represent the mixed configuration, which is suboptimal. Notice that they do not intersectas it can be clearly seen from the plot on the right, top line.
\label{plot geodesics}}
\end{figure}

For a two dimensional boundary theory, the region $A$ is simply a one dimensional segment of length $\ell$ at fixed boundary time $t$. The extremal surface we are looking for is given by the geodesic connecting the two extrema of this segment and extending in the bulk.
The Vaidya geometry (\ref{metric vadya d dim})  for $d=2$ interpolates between $AdS_3$ in Poincar\'e coordinates and the BTZ black hole of \cite{Banados:1992wn}.
By employing $r \equiv 1/z$ as holographic coordinate, the equations (\ref{cons eq}) and (\ref{eom d dim z}) become respectively \cite{Hubeny:2007xt, AbajoArrastia:2010yt}  
 \begin{equation}
 \label{vaidya 3dim cons}
\frac{r^4}{r_\ast^2} = -\big[ \, r^2-m(v)\, \big](v')^2  +2\,r' v' +r^2
\end{equation}
and
 \begin{equation}
  \label{vaidya 3dim eom}
r\,v'' - 2\,r' v'  + r^2\big[\, (v')^2-1\, \big]= 0\;.
\end{equation}
The first important feature of the three dimensional case is the kind of leading divergence in the expansion of the entanglement entropy $S_A$, which is not power like but logarithmical \cite{Callan:1994py, Holzhey:1994we, Calabrese:2004eu}. Roughly, this could be justified by observing that in the two dimensional boundary theory there is no  ``area law'' behavior because the $\partial A$ is made by two points and it has null measure. 
In three dimensions the regularized area (\ref{Ad reg def}) coincides with the regularized length of the geodesic 
 \begin{equation}
 \label{L2 reg def}
L_{2,\textrm{reg}} \,\equiv\, 
\lim_{\eta \,\rightarrow\,0^+} \left(
2 \int_{0}^{\frac{\ell}{2}-\eta}
\frac{z_\ast}{z^{2}}\,dx
+ 2\log \epsilon\right)
\hspace{1.6cm}
\epsilon\,\equiv\,z(\ell/2-\eta)
\end{equation}
where $\epsilon > 0$ is the UV cutoff of the holographic direction.
The limiting regimes at early and late times are respectively $AdS_3$ (bottom black curve in the figure \ref{plot Lregposkink}) and BTZ (top black curve), whose regularized lengths read \cite{Ryu:2006bv, Ryu:2006ef}
 \begin{equation}
  \label{Lreg ads3 btz}
L_{2,\textrm{reg}} \,\Big|_{AdS_{3}} =\, 2 \log \ell
\hspace{2cm}
L_{2,\textrm{reg}} \,\Big|_{\textrm{\tiny BTZ}} = \,
2\log\left[\frac{\beta_H}{\pi} \,\sinh\left(\frac{\pi \ell}{\beta_H}\right) \right]
\end{equation}
where $\beta_H \equiv 1/T_H = 2\pi/\sqrt{M}$ for $l=1$. 
By employing the well known Brown-Henneaux central charge $c=3l /(2G_N^{(3)})$ \cite{Brown:1986nw} for asymptotically $AdS_3$ spaces and the lengths (\ref{Lreg ads3 btz}) for the geodesics in $AdS_3$ and in the BTZ black hole, in \cite{Ryu:2006bv, Ryu:2006ef} it was checked that the holographic prescription (\ref{RT formula}) reproduces the expressions for the entanglement entropy of a single interval of length $\ell$ in a two dimensional CFT with central charge $c$ defined on an infinite line at $T=0$ and $T>0$ respectively \cite{Callan:1994py, Holzhey:1994we, Calabrese:2004eu}
 \begin{equation}
  \label{EE standard}
S_A\,\Big|_{T\,=\,0} = \, \frac{c}{3}\, \log\left(\frac{\ell}{\epsilon}\right)
\hspace{2cm}
S_A\,\Big|_{T\,>\,0} = \,\frac{c}{3} \,\log\left[\frac{\beta}{\pi \epsilon} \,\sinh\left(\frac{\pi \ell}{\beta}\right) \right]\;.
\end{equation}
Here $\beta\equiv 1/T$ and the second expression in (\ref{EE standard}) reduces to the first one when $\beta \rightarrow +\infty$, as expected. The same happens in (\ref{Lreg ads3 btz}) for $M \rightarrow 0$.\\
Besides $AdS_3$ and the BTZ black hole, there is another limiting regime of the Vaidya background in three dimensions where the holographic entanglement entropy has been computed analytically  \cite{Balasubramanian:2010ce,Balasubramanian:2011ur}: the  {\it infinitely thin shell limit}, defined when the thickness of the shell vanishes $a_v \rightarrow 0$. In this limit the metric is (\ref{metric vadya d dim}) with the mass profile given by a step function $m(v) = M \, \theta(v)$ at $v=0$, the non vanishing component $T_{vv}$ of the energy-momentum tensor is proportional to a delta function $\delta (v)$ and the infalling shell represents a shock wave.
The resulting geometry is given by a planar black brane (a BTZ black hole with a planar horizon) outside the shock wave and by $AdS_3$ inside the shock wave. 
The geodesics in both these regimes are known analytically. The geodesics entering inside the shell in the full shock wave geometry are piecewise curves made by merging the BTZ geodesic outside the shell with the $AdS_3$ geodesic inside.
In  \cite{Balasubramanian:2010ce,Balasubramanian:2011ur} a refraction law has been introduced which tells us how to merge them.
Once the merging point has been understood, the length of these geodesics is given by the sum of the three pieces: the first one going from one extremum of $A$ to the shell in the BTZ geometry, the second one inside the shell connecting two merging points of the shell at the same $z$ but different $x$ ($AdS_3$) and a third one going from this other point of the shell to the other extremum of $A$ on the boundary, again in the BTZ geometry (see the figure \ref{plot geodesics}). \\
Denoting by $r_{\textrm{\tiny H}}$ the position of the horizon of the BTZ geometry outside the shell, the regularized length of the geodesics entering into the shell can be written as \cite{Balasubramanian:2010ce,Balasubramanian:2011ur}
\begin{equation}
\label{Lreg thin shell}
L_{2,\textrm{reg}} \,\Big|_{\textrm{\tiny thin shell}}=\,2\, \log\left(\frac{\sinh(r_{\textrm{\tiny H}}\,  t)}{r_{\textrm{\tiny H}} \,s(\ell,t)}\right)
\end{equation}
where the function $s=s(\ell,t)\equiv \sin\theta \in [0,1]$ with $\theta \in [0,\pi/2]$ can be extracted from
\begin{equation}
\label{ldef thin shell}
\ell\,=\,\frac{1}{r_{\textrm{\tiny H}}}\left[\,\frac{2\cos\theta}{\rho \sin\theta}
+\log\left(\frac{2(1+\cos\theta)\rho^2+2\rho \sin\theta-\cos\theta}{2(1+\cos\theta)\rho^2-2\rho \sin\theta-\cos\theta}\right)\right]
\end{equation}
with $\rho$ defined as follows
\begin{equation}
\rho\,\equiv\,\frac{1}{2}\left( \coth(r_{\textrm{\tiny H}} \,t)+\sqrt{\coth^2(r_{\textrm{\tiny H}}\, t)-\frac{2\cos\theta}{1+\cos\theta}}\,\right)\;.
\end{equation}
In order to plot $L_{2,\textrm{reg}}$ as function of $\ell$ and of the boundary time $t$, we need both $L_{2,\textrm{reg}}|_{\textrm{\tiny BTZ}}$ (\ref{Lreg ads3 btz}) and $L_{2,\textrm{reg}}|_{\textrm{\tiny thin shell}}$ (\ref{Lreg thin shell}). Indeed, for small values of $\ell$ the geodesic is outside the shell and $L_{2,\textrm{reg}}$ is given by $L_{2,\textrm{reg}}|_{\textrm{\tiny BTZ}}$. As $\ell$ increases, at some point the geodesic enters into the shell and for $\ell$ larger than this value  $L_{2,\textrm{reg}}$ is given by  $L_{2,\textrm{reg}}|_{\textrm{\tiny thin shell}}$.  From (\ref{ldef thin shell}) one can observe that, for any fixed boundary time $t$, the function $\ell$ decreases as $\theta$ goes from $0$ to $\pi/2$. Thus, the critical value of $\ell$ from which we have to start using $L_{2,\textrm{reg}}|_{\textrm{\tiny thin shell}}$ is 
\begin{equation}
\label{ell critical}
\ell\, \big|_{\theta =\pi/2}
\,=\,\frac{1}{r_{\textrm{\tiny H}}}\,\log\left(\frac{\rho \,|_{\theta=\pi/2}+1}{\rho \,|_{\theta=\pi/2}-1}\right)
\,=\,\frac{1}{r_{\textrm{\tiny H}}}\,\log\left(\frac{\coth(r_{\textrm{\tiny H}}\, t)+1}{\coth(r_{\textrm{\tiny H}}\, t)-1}\right)
\,=\, 2t\;.
\end{equation}
Given a boundary time $t$, this is the value of $\ell$ after which the geodesic enters into the shell.
In the figure \ref{plot Lregposkink} the $a_v=0$ limit we are discussing is represented by the dashed curves. The relation (\ref{ell critical}) can be checked on those plots, since the dashed curve characterized by $t$ deviates from the BTZ continuous black curve at $\ell=2t$. This is the relation found in \cite{Calabrese:2005in} for two dimensional CFT models between the duration of the linear increasing of the entanglement entropy after a global quench and the size $\ell$ of the spatial interval $A$, which lead the authors to suggest the quasiparticles picture.

\section{Holographic mutual information}
\label{section HMI}

In this section we consider the holographic mutual information for three and four dimensional Vaidya metrics, employing also the analytic solution for the thin shell limit of the three dimensional case. We numerically compute its dependence from the size and the distance of the strips and also study the transition curves in the configuration space, finding that a time independent region exists where the holographic mutual information vanishes at all times.

When the boundary region $A$ we are interested in is made by two disjoint regions $A = A_1 \cup A_2$ with $A_1 \cap A_2 = \emptyset$, the situation becomes more complicated and also more interesting.
The most important quantity to consider in this case is the mutual information
\begin{equation}
\label{MI def}
I(A_1,A_2)\,\equiv\,S_{A_1} + S_{A_2} - S_{A_1 \cup A_2}  
\end{equation}
because in this linear combination the leading divergence due to the area law is canceled. The quantity $I(A_1,A_2)$ measures the classical and quantum correlation between $A_1$ and $A_2$. \\
Within the context of two dimensional CFT, the Renyi entropies $S^{(n)}_{A_1 \cup A_2}  \equiv (1-n)^{-1}\log \textrm{Tr} \rho^n_{ A_1 \cup A_2}$ (for interger $n\geqslant 2$) have been computed for  the free compactified boson \cite{Calabrese:2009ez} and for the Ising model \cite{Calabrese:2010he}. These results have been checked against existing numerical data on spin chains \cite{Furukawa:2008uk, Fagotti:2010yr}. Unfortunately, the analytic continuation of these quantities for $n \rightarrow 1$, which is needed to find the mutual information through derivation $S_{A_1 \cup A_2}  = - \,\partial_n \textrm{Tr} \rho^n_{ A_1 \cup A_2}\big|_{n=1}$, is still unknown for the general expressions, but it has been done for some limiting regimes of the parameters like the decompactification regime for the free boson (when the field takes values on the whole real line) \cite{Calabrese:2009ez} or when the two intervals are very far apart \cite{Calabrese:2010he}. The main quantitative lesson one learns from these results is that the Renyi entropies, or equivalently the Renyi mutual information $I^{(n)}(A_1,A_2)$, which is defined in the obvious way by combining the Renyi entropies as done in (\ref{MI def}) for the mutual information, encodes all the data of the CFT. In other words, $\textrm{Tr} \rho^n_{ A_1 \cup A_2}$ do not contain only the central charge $c$, like $S_A$ for one interval, but also all the conformal dimensions and all the OPE coefficients of the theory.
\begin{figure}[th]
\vspace{.4cm}
\begin{tabular}{ccc}
\hspace{-.8cm}
\includegraphics[width=3.2in]{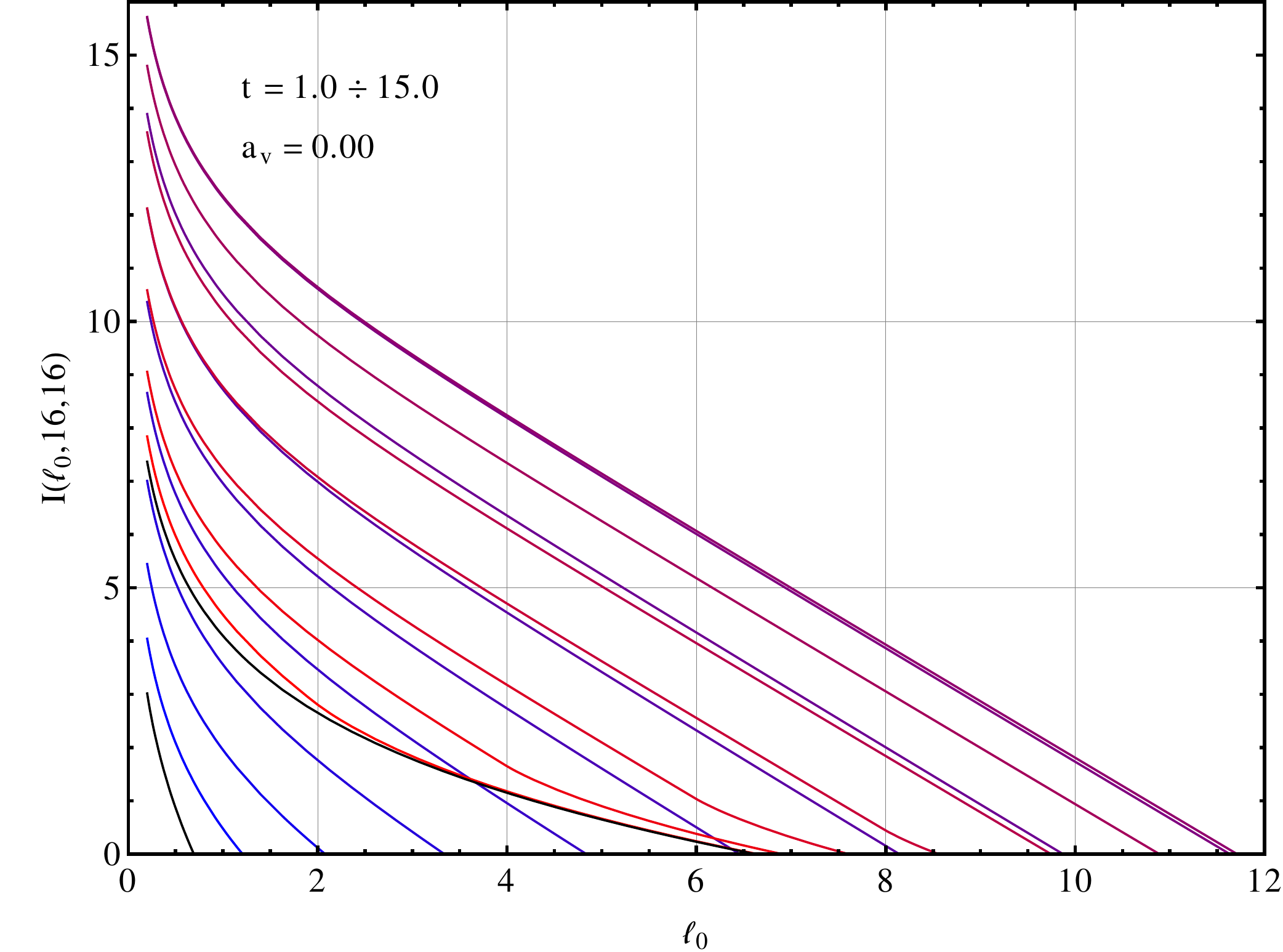}
& \hspace{-.7cm} & 
\includegraphics[width=3.2in]{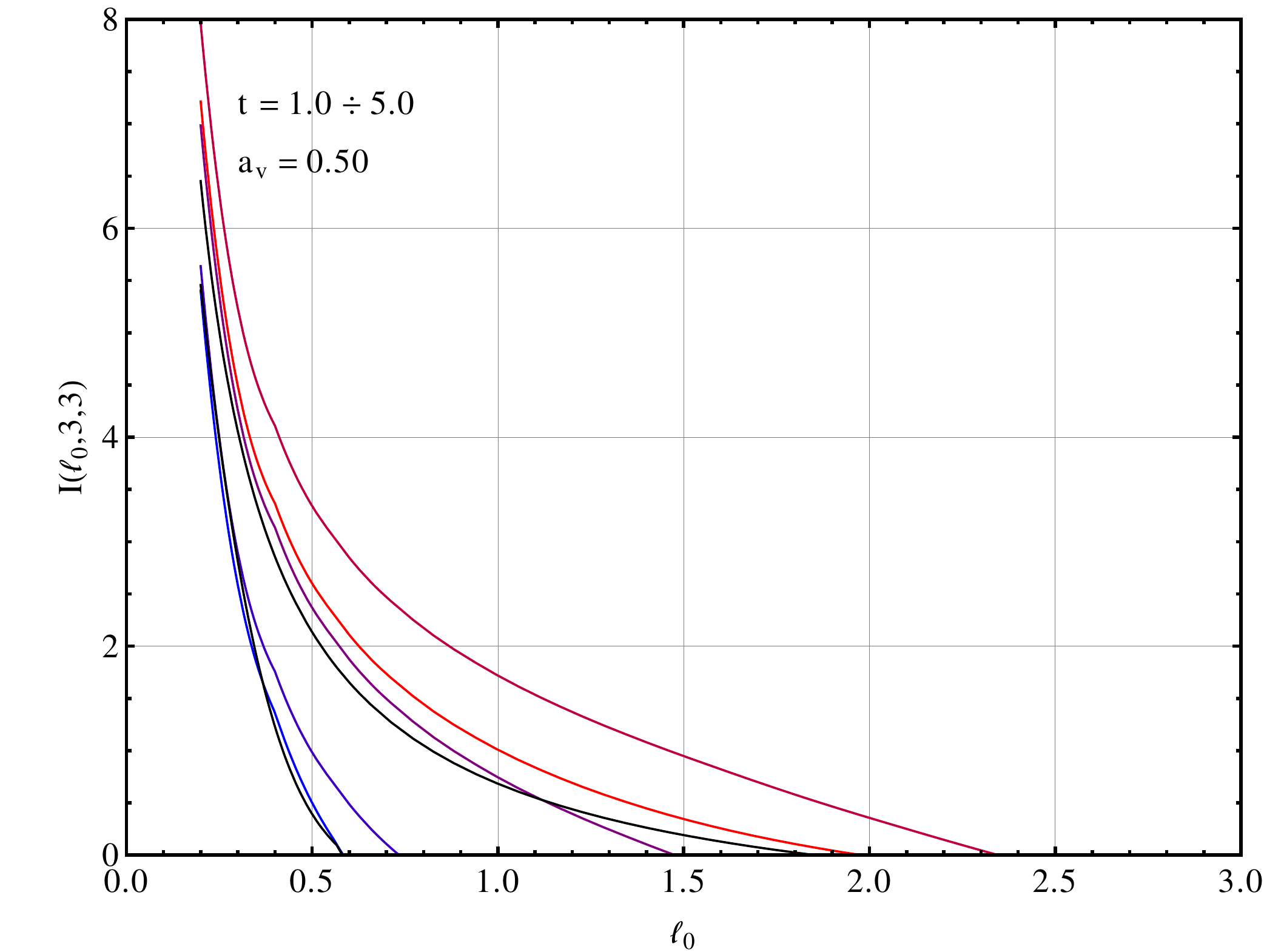}
\end{tabular}
\caption{Holographic mutual information $I(\ell_0, \ell_1, \ell_1)$ in terms of $\ell_0$ for Vaidya metrics in three (plots on the left, infinitely thin shell regime) and four dimensions (plots on the right) in the bulk. Different curves are characterized by the boundary time $t$, whose value increases going from the red curves to the blue ones with $\Delta t =1$ and within the range indicated.
The black curves correspond to $AdS_{d+1}$ (top curve) and Schwarzschild black hole (bottom curve). The transition of the holographic mutual information is continuos with a discontinuous first derivative.
\label{plot MI L0 dep}}
\end{figure}

From the AdS/CFT point of view, it is natural to study the holographic mutual information, which is defined like in (\ref{MI def}) with $S_A$ given by the holographic formula (\ref{RT formula}), as a further test of the holographic prescription (\ref{RT formula}) for the entanglement entropy. It is known \cite{Headrick:2010zt, Tonni:2010pv} that the holographic mutual information displays a continuos transition from zero to positive values with a discontinuous first derivative which is not observed  in the simple CFT models considered so far \cite{Calabrese:2009ez, Calabrese:2010he}. This feature is believed to be a large $c$ effect.\\
Given the two disconnected regions $A_1$ and $A_2$ in the boundary, there are three configurations of two surfaces extending in the bulk whose boundaries coincide with $\partial A =\partial A_1 \cup \partial A_2$: the first is simply the union of the two surfaces characterizing $S_{A_1}$ and $S_{A_2}$ (which bounds two disconnected volumes and hence it will be referred to as ``disconnected'' configuration). The second one is the  ``connected''  configuration, given by a bridge connecting $A_1$ and $A_2$ through the bulk (bounding a single connected volume in the bulk). 
The third configuration is composed by the extremal surface connecting the first extrema of the regions (let us assume for the moment we have strips) and the one connecting the second ones. We will denote this case as ``mixed''  configuration. In this mixed configuration, for a static background, the extremal surfaces intersect, but for a dynamical background this is not generally true (see the set of green geodesics in the figure \ref{plot geodesics}). Also, for two disjoint circular regions in a 2+1 dimensional boundary, this type configuration does not occur at all.
A similar mixed configurations of surfaces also occurs in the proof of the strong subadditivity for the holographic entanglement entropy \cite{Headrick:2007km}. Since they do not always intersect for time dependent backgrounds, the proof cannot be extended to the dynamical case in a straightforward way. In our case, since $\mathcal{A}_{d,\textrm{reg}}$ is an increasing function of the size of $A$ at any fixed time $t$ (see the figures \ref{plot Lregposkink} and \ref{plot Lregposkink d=3})
we can claim that this mixed configuration is always suboptimal with respect to the disconnected one (see also the upper line of the figure \ref{plot intersecting configs}).
Thus, the regularized area entering in the holographic computation of $S_{A_1 \cup A_2}$  reads
\begin{equation}
\label{Adreg comparison}
\mathcal{A}_{d,\textrm{reg}} \,= \,\textrm{min}\Big(
\mathcal{A}_{d,\textrm{reg}}\big|_{\textrm{connected}}\,  , 
\mathcal{A}_{d,\textrm{reg}}\big|_{\textrm{disconnected}}
\Big)\,.
\end{equation}
Notice that both configurations display the same UV divergence which cancels in the linear combination (\ref{MI def}). In three dimensional backgrounds ($d=2$), the lengths of the geodesics are involved in (\ref{Adreg comparison}).\\ 
It is straightforward to notice that when the disconnected configuration is the minimal one in (\ref{Adreg comparison}) the holographic mutual information is zero, while it is positive in the other case. Moreover, when $A_1$ and $A_2$ are very far apart from each other the disconnected configuration is  minimal. Thus, considering a configuration space which describes all the possible sizes and the relative distance between $A_1$ and $A_2$, there must be some region of this space where the ``connected'' configuration is minimal  and some other region where, instead, the ``disconnected'' configuration is minimal. The corresponding holographic mutual information is zero and positive respectively. The curve in the configuration space which characterizes this transition in the configuration space is given by the following equation
\begin{equation}
\label{transition eq general}
\mathcal{A}_{d,\textrm{reg}}\big|_{\textrm{connected}}
\,=\,
\mathcal{A}_{d,\textrm{reg}}\big|_{\textrm{disconnected}}\;.
\end{equation}
This equation is not easy to solve and must be studied case by case. The examples of $AdS_{d+1}$ and of the charged black hole in four dimensions have been considered in \cite{Headrick:2010zt, Tonni:2010pv}. Here we study this equation for the Vaidya metrics (\ref{metric vadya d dim}) in three and four dimensions.

\begin{figure}[th]
\vspace{.4cm}
\begin{tabular}{ccc}
\hspace{-.8cm}
\includegraphics[width=3.2in]{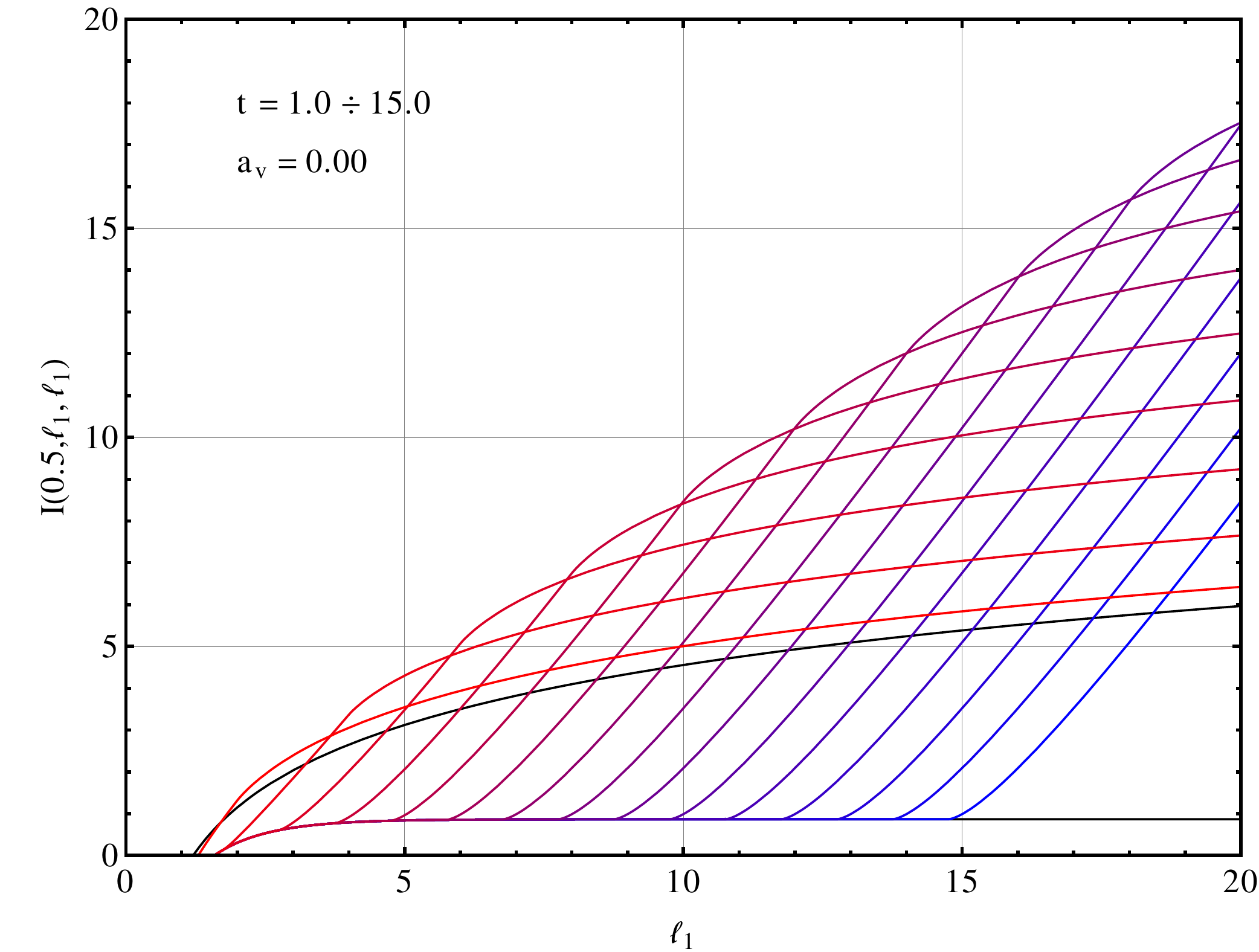}
& \hspace{-.8cm} & 
\includegraphics[width=3.2in]{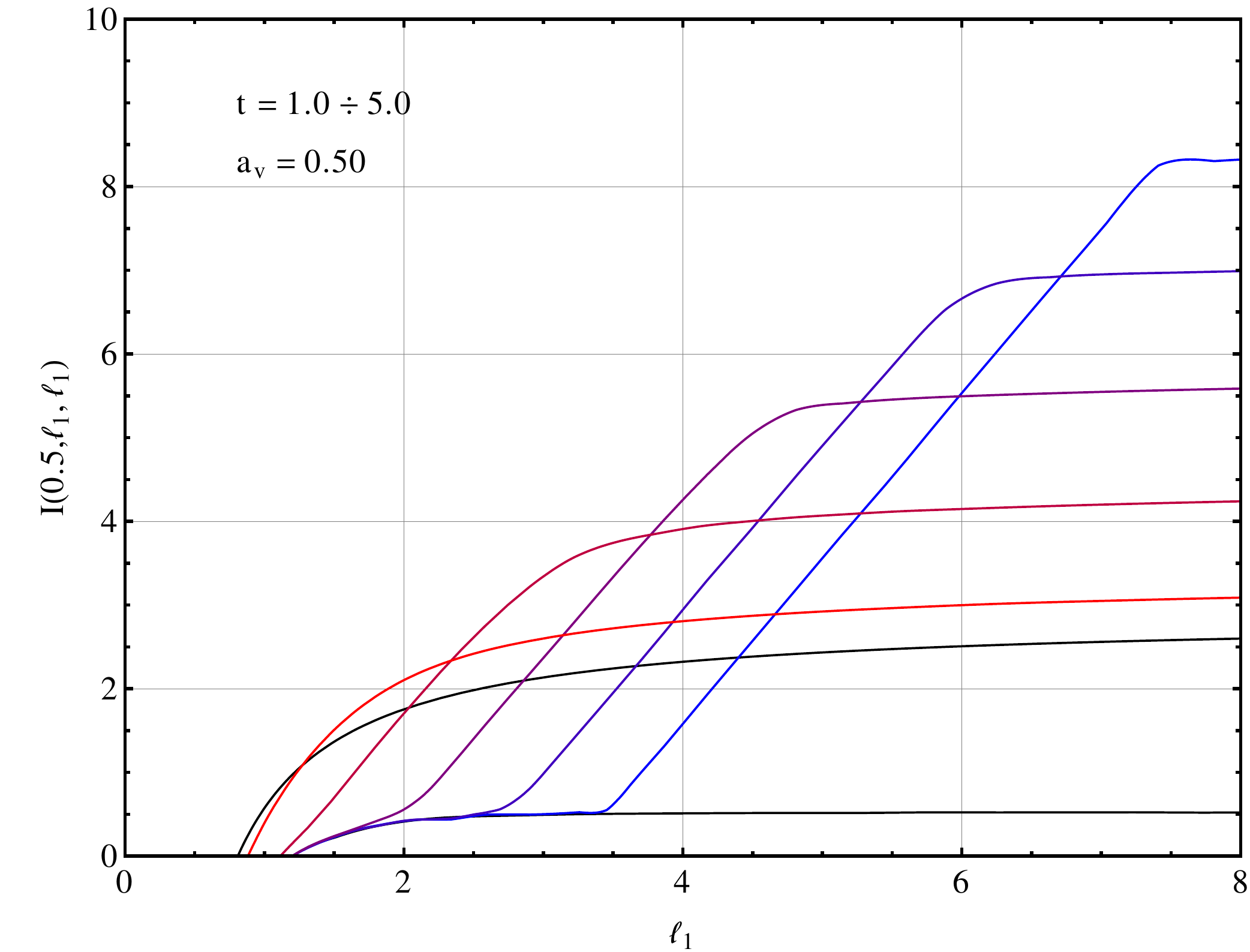}
\end{tabular}
\caption{Holographic mutual information $I(\ell_0, \ell_1, \ell_1)$ as function of $\ell_1$ at fixed $\ell_0$ for Vaidya metrics in three (plots on the left, infinitely thin shell regime) and four dimensions (plots on the right) in the bulk. The other parameters are the same ones employed in the figure \ref{plot MI L0 dep}.
\label{plot MI L1 dep}}
\end{figure}

Let us first consider the holographic mutual information $I(\ell_0, \ell_1, \ell_2)$ (see (\ref{MI def})) for the Vaidya metrics (\ref{metric vadya d dim}) in three and four dimensions with the mass profile given by (\ref{mass pos kink}). This quantity depends on many variables and our analysis is mainly numerical. We take equal strips $\ell_2=\ell_1$ for our plots unless indicated otherwise. \\
In the figures \ref{plot MI L0 dep}, \ref{plot MI L1 dep} we show the dependence of $I(\ell_0, \ell_1, \ell_1)$ from the distance $\ell_0$ between the intervals and the size $\ell_1$ respectively. 
The continuos black curves represent the corresponding quantities in the limiting regimes of $AdS_{d+1}$ and Schwarzschild black hole (BTZ black hole in three dimensions), while the colored ones are characterized by intermediate boundary times indicated in the plots. 
Notice that the qualitative features of the curve do not change with the number of dimensions. In general we can clearly observe the transition of the mutual information from positive values to zero when $\ell_0$ increases at fixed size $\ell_1$ (figure \ref{plot MI L0 dep}) and from zero to positive values as $\ell_1$ increases at fixed separation $\ell_0$ (figure \ref{plot MI L1 dep}). This transition is continuos with a discontinuous first derivative and unfortunately we do have a clear understanding of it. We recall that the holographic mutual information is positive when the connected configuration is favored. 
When we plot a family of curves parameterized by the boundary time $t$, the common feature one observes is that the bigger $t$ is, the larger is the range of variables where the curve characterized by $t$ reproduces the corresponding black hole result. Then, for any finite $t$ at some point the curve deviates from the black hole behavior and tends asymptotically to the $AdS_{d+1}$ behavior, eventually shifted by a constant.\\
In three dimensions ($d=2$) we can take advantage from the fact that the exact solution is known of the thin shell limit case $a_v \rightarrow 0$ (see (\ref{Lreg thin shell})) \cite{Balasubramanian:2010ce,Balasubramanian:2011ur} and working with the analytic solution allows us to extend the range of the variables we can explore.

\begin{figure}[th]
\begin{tabular}{ccc}
\hspace{-.8cm}
\includegraphics[width=3.2in]{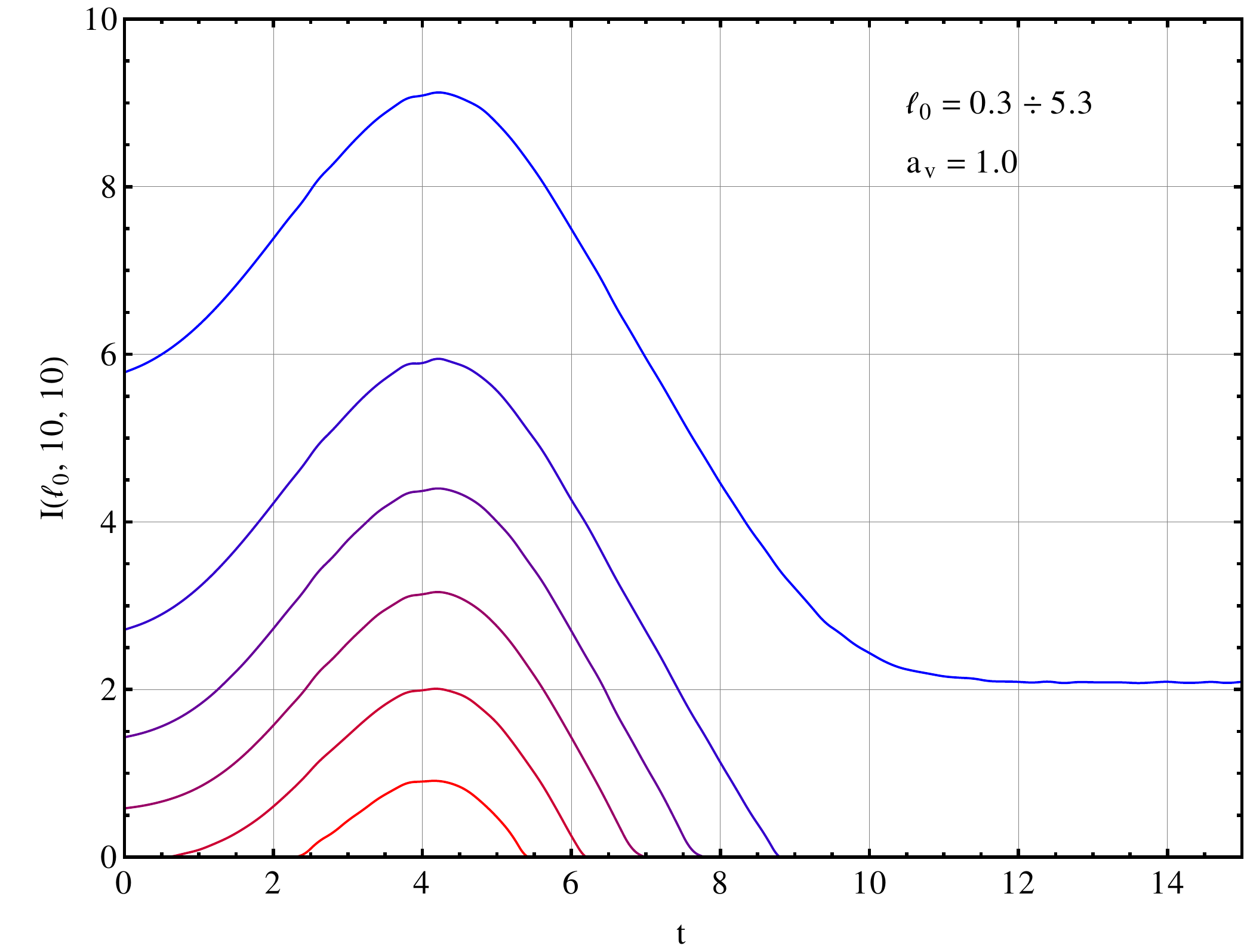}
& \hspace{-.7cm} & 
\includegraphics[width=3.2in]{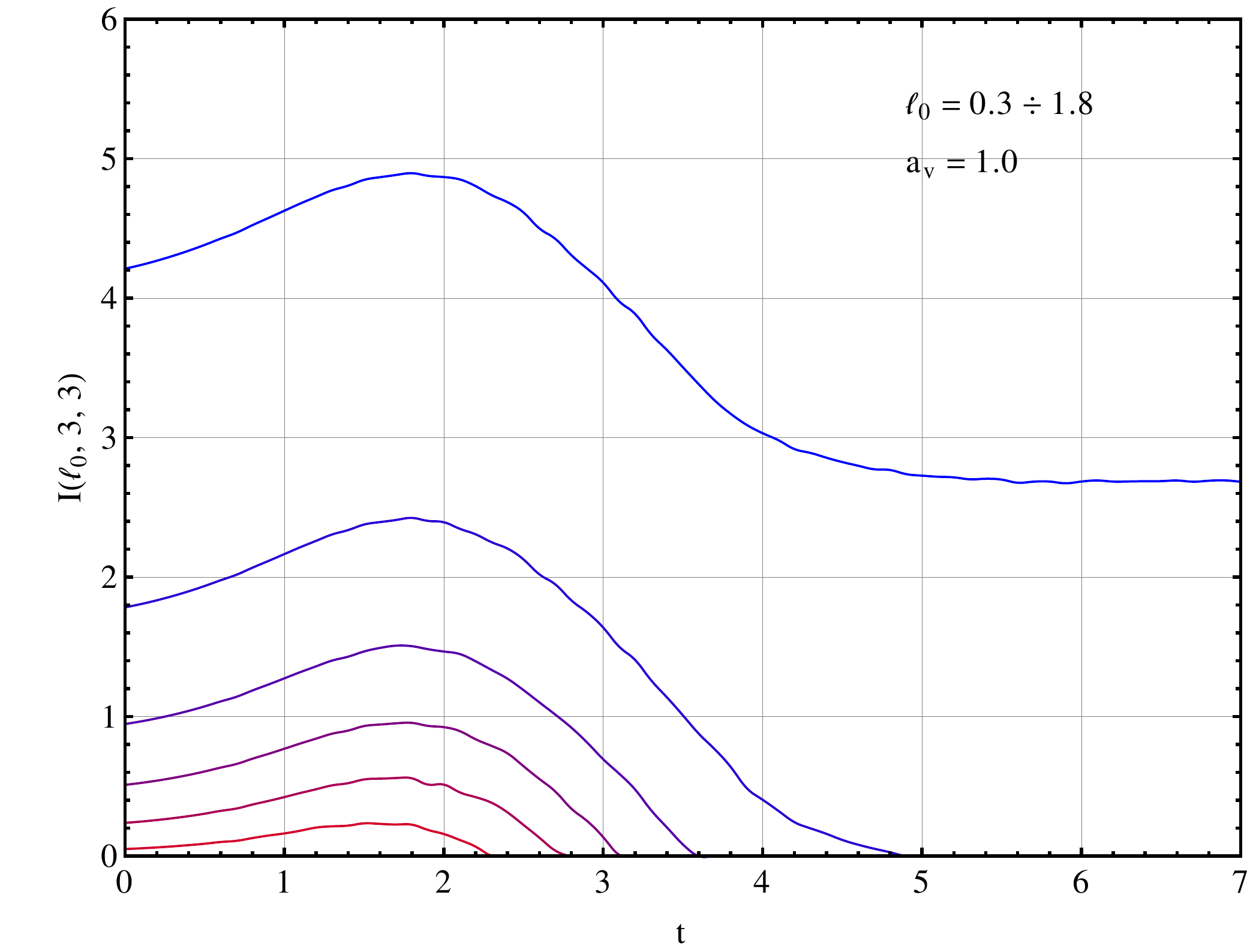}
\end{tabular}
\caption{Holographic mutual information $I(\ell_0, \ell_1, \ell_1)$ as function of the boundary time $t$ at fixed $\ell_1$. The different curves are characterized by different values of $\ell_0$ which increase going from the top curve to the bottom one. As in the previous figures, we show the $d=2$ case on the left and the $d=3$ case on the right. For a fixed value of $\ell_1$, varying  $\ell_0$ three different behaviors are observed.
\label{plot MI t dep}}
\end{figure}

As for the dependence on the boundary time $t$ of the holographic mutual information $I(\ell_0, \ell_1, \ell_1)$, this is shown in the figure \ref{plot MI t dep} for different values of the separation $\ell_0$ between the two strips and a fixed value of their size $\ell_1$ in three dimensions (plot on the left) and in four dimensions (plot on the right). From these plots we observe that, at a fixed of $\ell_1$, varying the separation $\ell_0$ between the strips four different behaviors are observed. For $\ell_0$ very large  
$I(\ell_0, \ell_1, \ell_1)$ is zero at all times. Decreasing $\ell_0$ (i.e. going from the bottom curves to the top ones in the figure \ref{plot MI t dep}) we find that $I(\ell_0, \ell_1, \ell_1)$ is zero at $t=0$, then it becomes positive for a finite range of $t$ and then it vanishes again. Decreasing further $\ell_0$ the holographic mutual information starts positive at $t=0$ but then it vanishes at some time. Then, for small $\ell_0$ we get that $I(\ell_0, \ell_1, \ell_1)$ is positive for any boundary time $t$, namely the connected configuration is always favored.

\begin{figure}[th]
\begin{tabular}{ccc}
\hspace{-.8cm}
\includegraphics[width=3.2in]{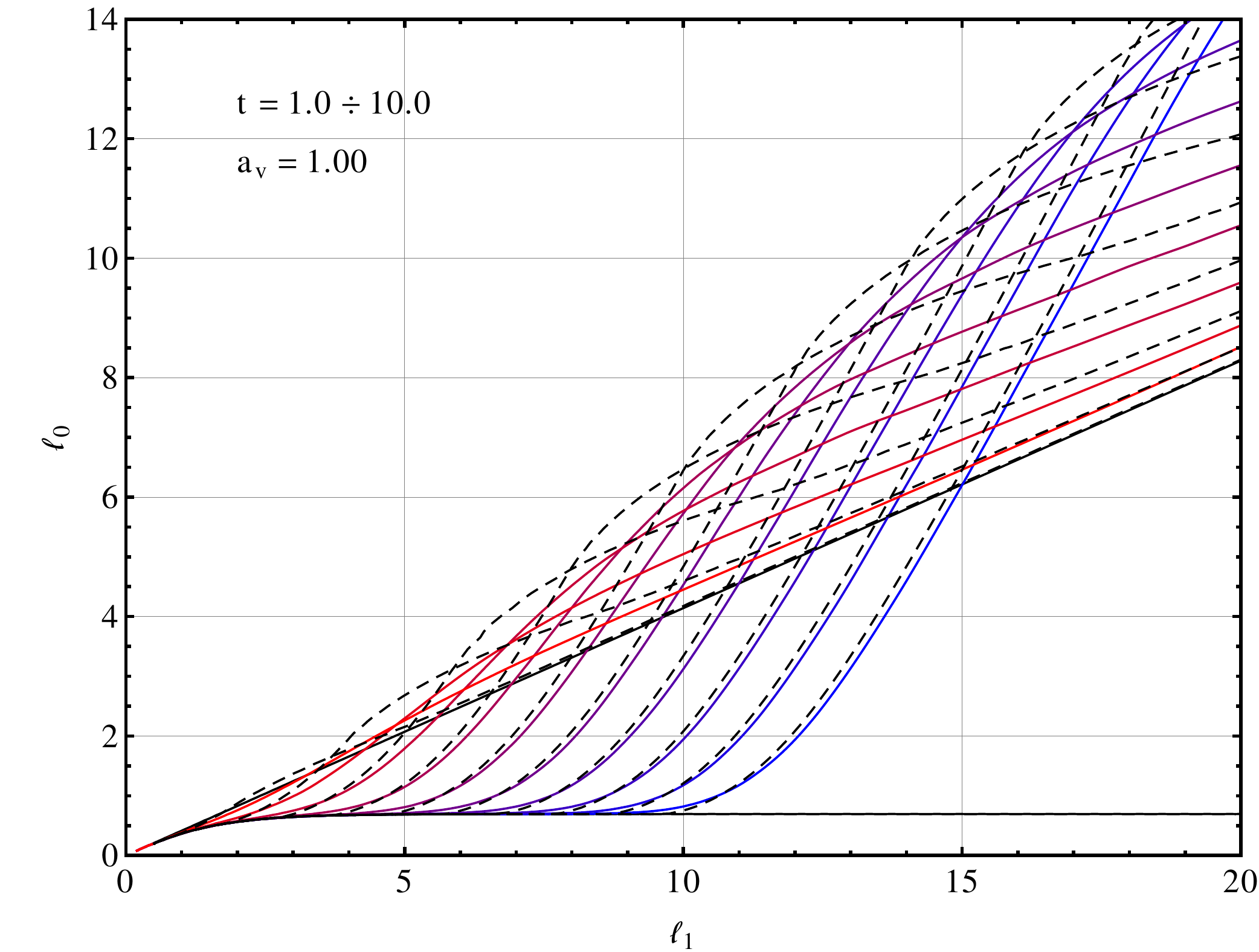}
& \hspace{-.7cm} & 
\includegraphics[width=3.2in]{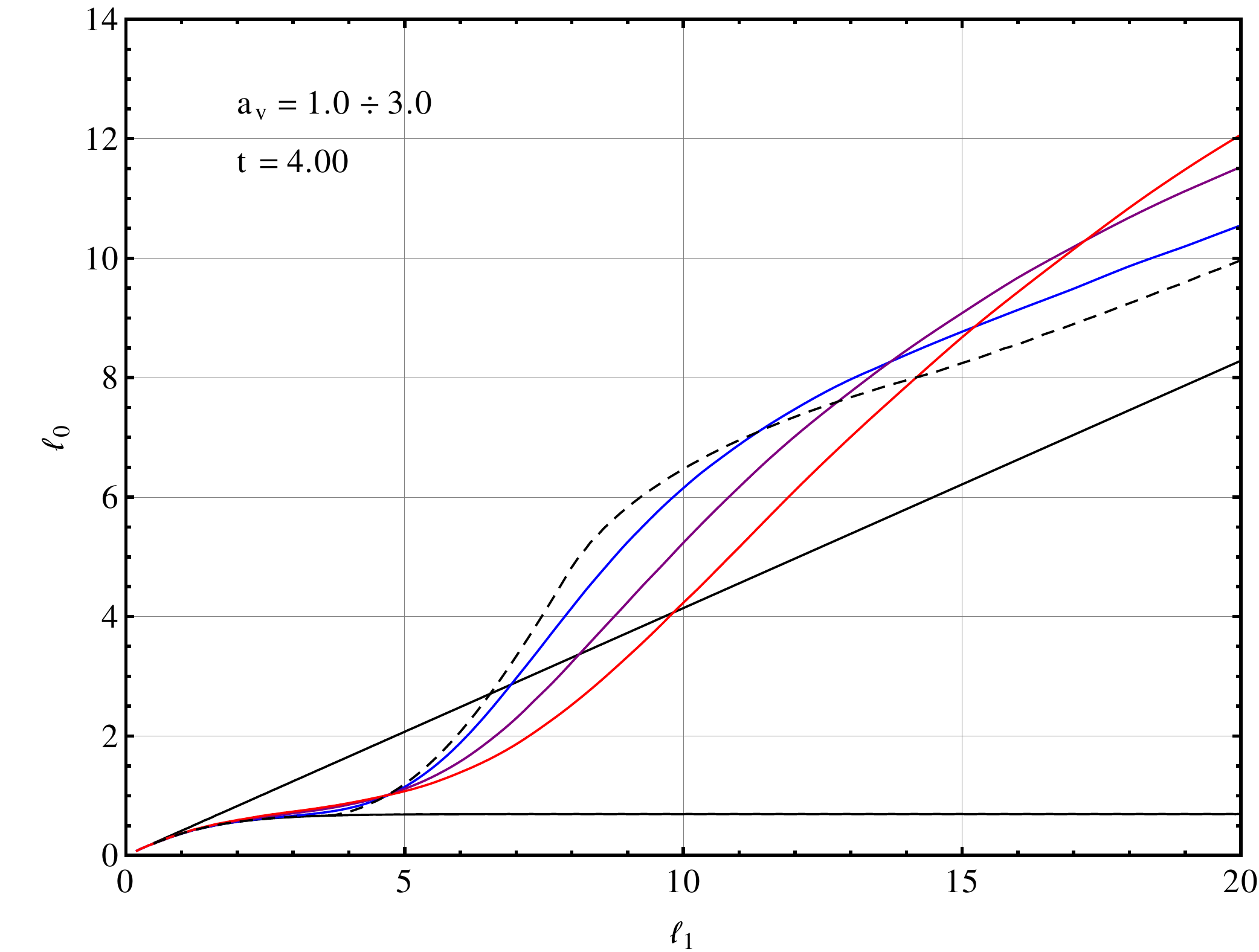}
\end{tabular}
\caption{Transition point of the holographic mutual information $I(\ell_0, \ell_1, \ell_1)$ for the three dimensional Vaidya metric in the configuration space $(\ell_1,\ell_0)$. The black curves represent the limiting regimes of $AdS_3$ (top curve, given by (\ref{ads3 transition curve}) at $\omega=1$) and BTZ (bottom curve, given by (\ref{btz transition curve}) at $\omega=1$).
On the left we plot the transition point for different times (increasing as we go from the red curve to the blue one) and thickness $a_v=1$ (the dashed curve represent the thin shell limit $a_v=0$). On the right, the transition point in the configuration space is plotted at a fixed time $t=4$ for various values of the thickness $a_v$ in (\ref{mass pos kink}): from $a_v=0$ (dashed curve) to $a_v=3$ (red curve). There is a whole region of the configuration space where the holographic mutual information is zero for any boundary time. 
\label{plot transposkink}}
\end{figure}

\begin{figure}[th]
\begin{tabular}{ccc}
\hspace{-.8cm}
\includegraphics[width=3.2in]{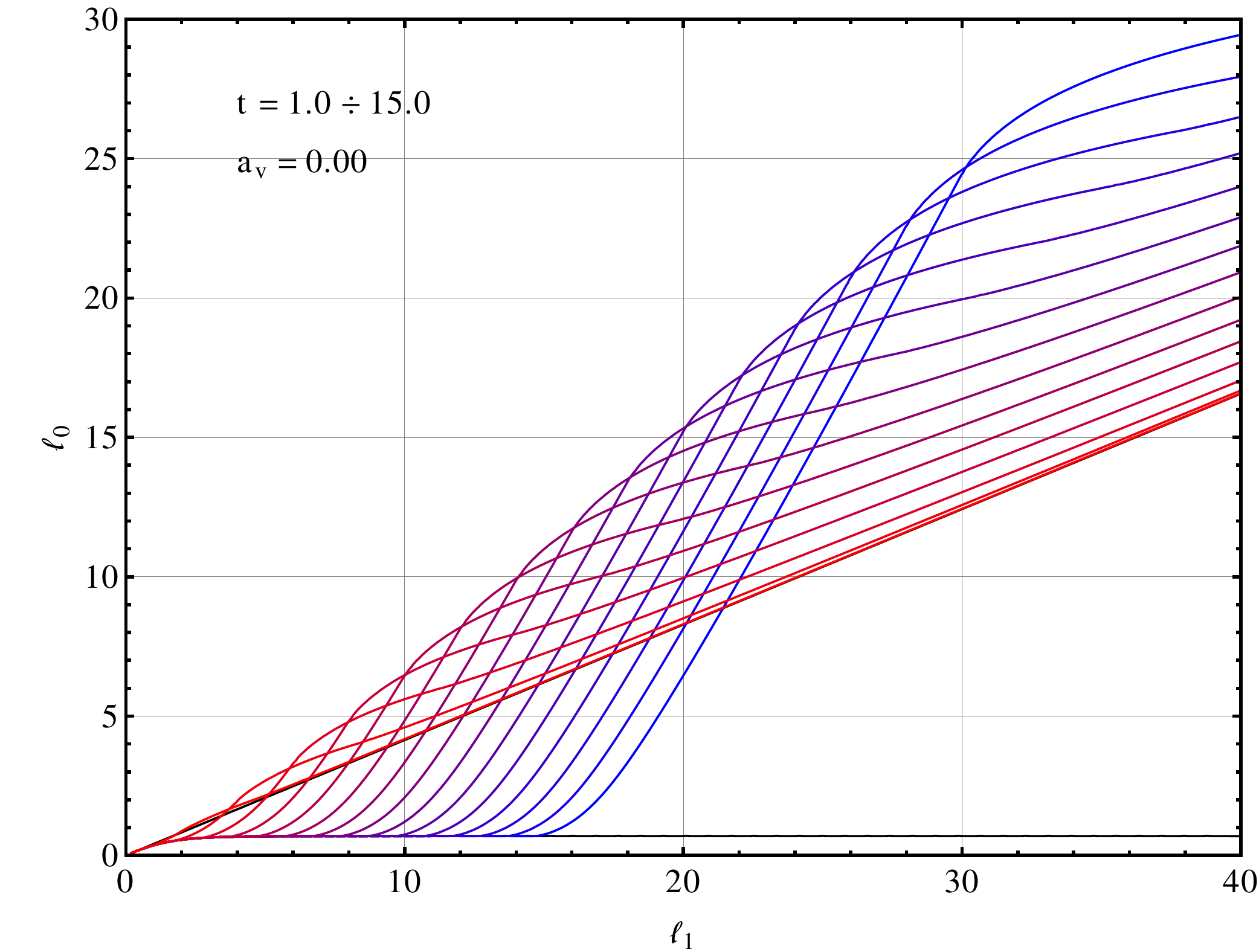}
& \hspace{-.7cm} & 
\includegraphics[width=3.2in]{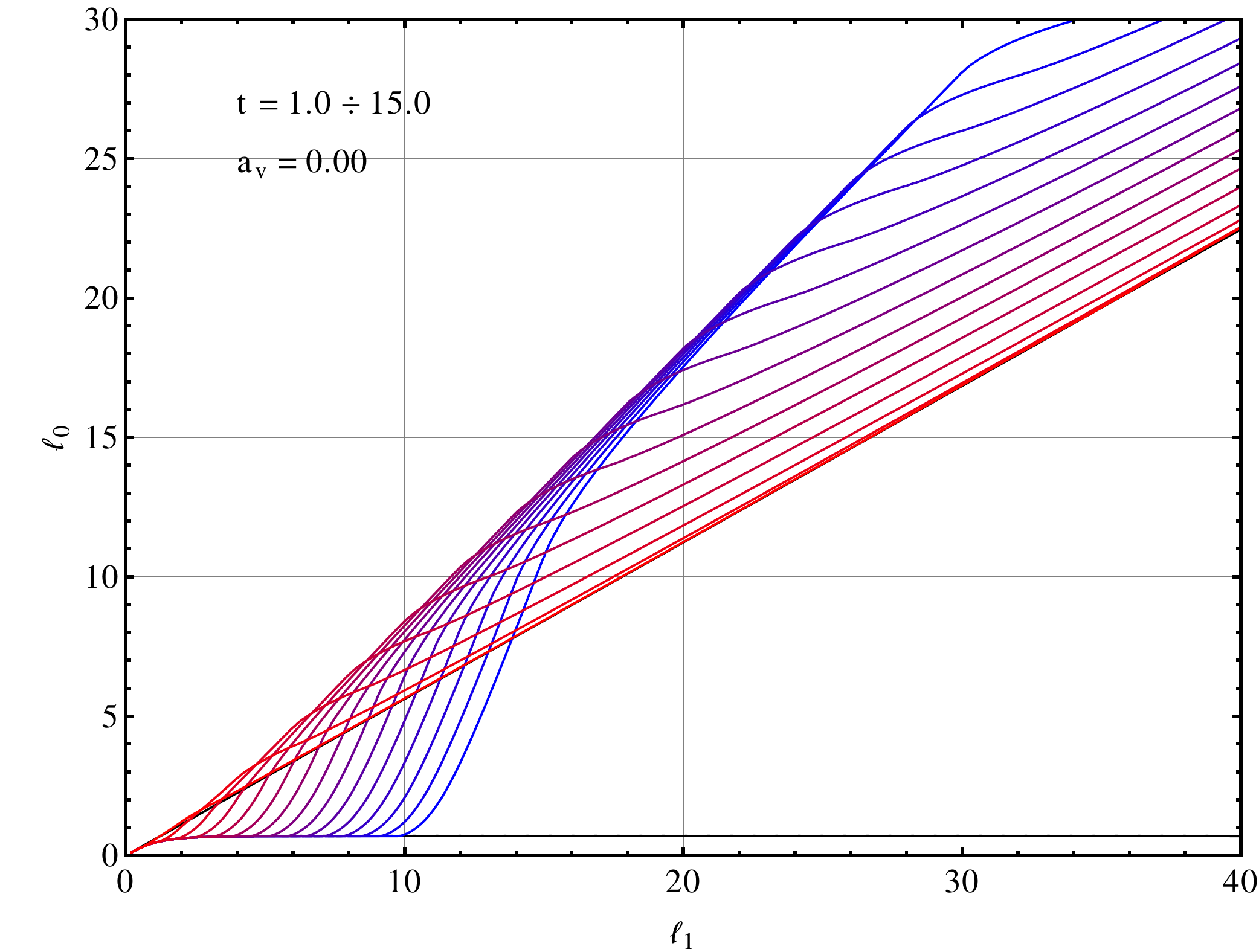}
\end{tabular}
\caption{Transition point of the holographic mutual information $I(\ell_0, \ell_1, \omega \ell_1)$ in the configuration space $(\ell_1,\ell_0)$ for the three dimensional Vaidya metric in the thin shell limit, whose analytic solution allows to extend the range of the configuration (see also the figure \ref{plot transposkink}). We give the curves for different boundary times which increase from the red curve to the blue one. On the left we set $\omega=1$ while on the right $\omega=2$. The curve above which the holographic mutual information vanishes for any boundary time depends on $\omega$.
\label{plot transition point thin shell}}
\end{figure}

\begin{figure}[th]
\begin{center}
\includegraphics[width=3.6in]{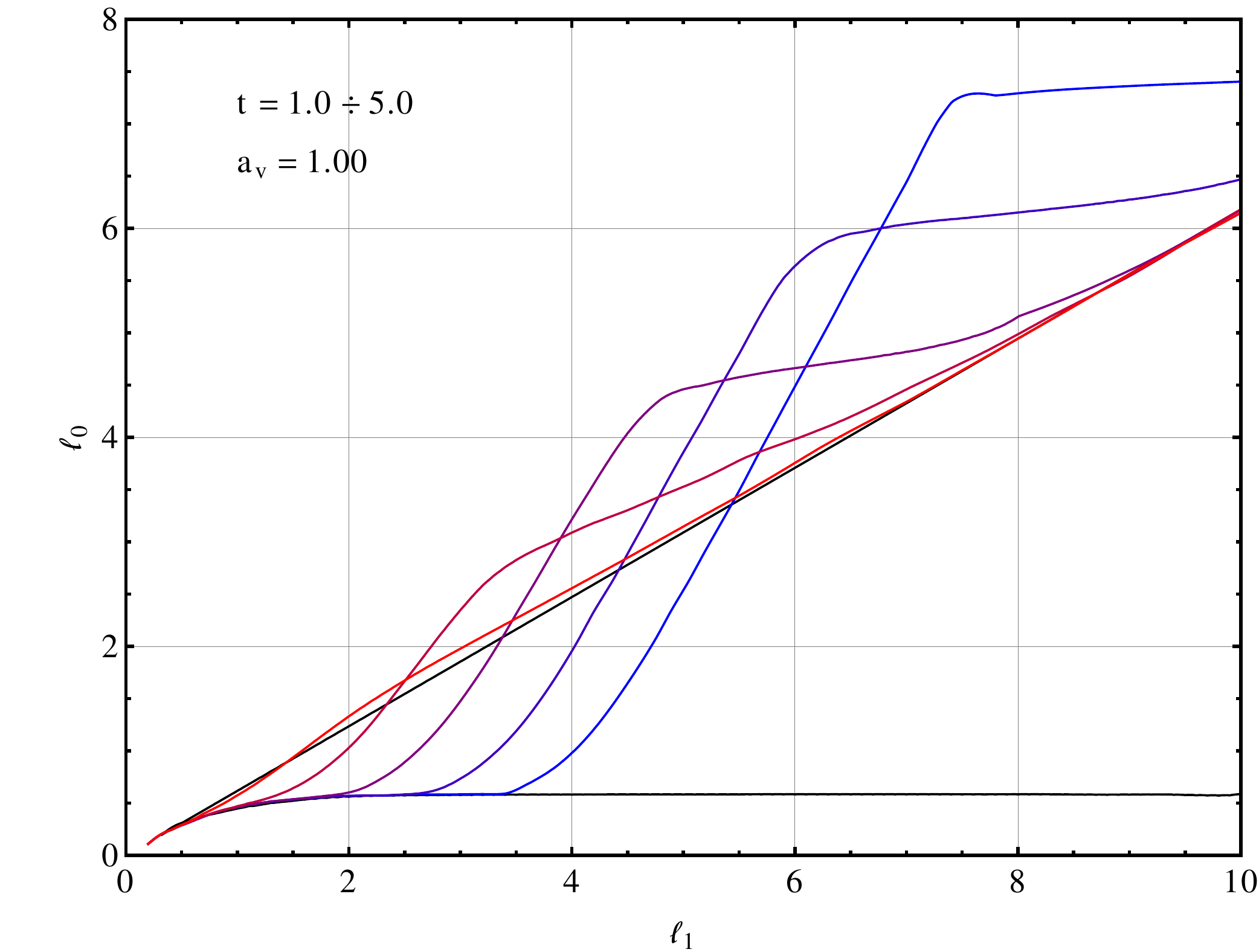}
\end{center}
\vspace{-.7cm}
\caption{Transition point of the holographic mutual information $I(\ell_0, \ell_1, \ell_1)$ for the four dimensional Vaidya metric in the configuration space $(\ell_1,\ell_0)$ at different boundary times which increase going from the red curve to the blue one. The black curves represent the limiting regimes of $AdS_4$ (top curve) and Schwarzschild black hole (bottom curve). Also in four dimensions there is a curve above which the holographic mutual information is zero for any boundary time. 
\label{plot transition 4dim}}
\end{figure}

\noindent 
In order to describe these four regimes from another point of view, we find it useful to study the transition curve of $I(\ell_0, \ell_1, \ell_2)$ in the configuration space given by $\ell_2$, $\ell_1$ and $\ell_0$.
This means to find a family of curves parameterized by $t$ which solves the equation (\ref{transition eq general}). These transition curves are shown in the figures \ref{plot transposkink}, \ref{plot transition point thin shell} (three dimensional case) and \ref{plot transition 4dim} (four dimensional case).\\
We set $\ell_2 =\omega \ell_1$ for some finite $\omega>0$ and then consider the space $(\ell_1 ,\ell_0)$. In the figure \ref{plot transposkink} we study the three dimensional background for some $a_v>0$ and $\omega=1$. In the plot on the right the $a_v$ dependence is show at fixed $t$. 
The dashed curves correspond to the $a_v \rightarrow 0$ limit and are obtained through the analytic solution of \cite{Balasubramanian:2010ce, Balasubramanian:2011ur}.
The main feature we notice is that it is possible to draw a critical curve $\hat{\ell}_0(\ell_1)$ which is {\it independent of the time $t$} such that for any configuration specified by a point above this curve the holographic mutual information is zero at all times.
This critical curve is above the transition curve of $AdS_3$ and it depends on $a_v$. The region below it and above the $AdS_3$ transition curve becomes larger as $a_v$ becomes smaller, as it can be observed from the plot on the right in the figure \ref{plot transposkink}.
In the figure \ref{plot transition point thin shell} we consider this transition curves for the analytic solution of the $a_v \rightarrow 0$ limit in order to extend the range of the configuration space $(\ell_1,\ell_0)$ and to study the dependence on $\omega$, which is equal to 1 as usual on the left and equal to 2 on the right. 
Comparing the two plots we can clearly see the linear behavior of the critical curve for large $\ell_1$ and notice that it depends also on $\omega$. \\
In the figure \ref{plot transition 4dim} we consider the four dimensional background for $\omega=1$: the qualitative features are the same described above but the configuration space we can explore is much smaller because numerically we find it difficult to go further.

\subsection{Limiting regimes in three dimensions}

Here we discuss the analytic expressions of the transition curves for the
limiting regimes of the black hole formation process in three dimensions, namely $AdS_3$ for early times and the BTZ black hole in the late times (black curves in the figures \ref{plot MI L0 dep}, \ref{plot MI L1 dep}, \ref{plot transposkink} and \ref{plot transition point thin shell}).

The boundary theory is two dimensional and the two spatial regions $A_1$ and $A_2$ at $t=\textrm{const}$ are intervals whose lengths are respectively  $\ell_1= x_{21}$ and $\ell_2 =x_{43}$ (we adopt the notation $x_{ij} \equiv x_i - x_j$). The separation  length is $\ell_0 =x_{32}$ and therefore $\ell_1+\ell_2+\ell_0 = x_{41}$.

The transition of the holographic mutual information for $AdS_3$ has been studied in \cite{Headrick:2010zt}.
Introducing the harmonic ratio for the four extrema of the two intervals
\begin{equation}
\label{harmonic ratio def}
x\equiv \frac{x_{12}\, x_{34}}{x_{13}\, x_{24}} \,=\,\frac{\ell_1 \ell_2}{(\ell_1+\ell_0)(\ell_2+\ell_0)}
\end{equation}
and employing the first formula in (\ref{Lreg ads3 btz}), the holographic mutual information of two disjoint intervals for $AdS_3$ reads \cite{Headrick:2010zt}
\begin{equation}
\label{MI transition ads3}
I(A_1,A_2)
\,=\,
\left\{\begin{array}{lcl}
\displaystyle 0 & \hspace{.5cm}& x < 1/2\hspace{.2cm} \\
\rule{0pt}{.7cm}
\displaystyle
\frac{c}{3}\, \log\bigg(\frac{\ell_1 \ell_2}{\ell_0(\ell_1+\ell_2+\ell_0)}\bigg)
= \frac{c}{3}\, \log\bigg(\frac{x}{1-x}\bigg) 
&  & x >1/2
\end{array}\right.
\end{equation}
where we recall that $c=3l/(2G_N^{(3)})$ \cite{Brown:1986nw}.
The transition point $x=1/2$ corresponds to the solution of $\ell_1 \ell_2/[\ell_0(\ell_1+\ell_2+\ell_0)]=1$, namely when the argument of the logarithm in (\ref{MI transition ads3}) is equal to 1. 
Parameterizing $\ell_2$ as $\ell_2= \omega \ell_1$ with $\omega>0$, the equation $x=1/2$ becomes a second order equation for $\ell_0$ (see (\ref{harmonic ratio def})) with only one positive solution
\begin{equation}
\label{ads3 transition curve}
\ell_0 = \frac{1+\omega}{2}\left(\sqrt{\frac{4\omega}{(1+\omega)^2}+1}-1\right)\ell_1\;.
\end{equation}
This curve is a line in the plane $(\ell_1, \ell_0)$ passing through the origin whose angular coefficient depends on $\omega$. For $\omega=1$ it becomes $\ell_0 = (\sqrt{2}-1)\ell_1$ and this case is employed in the figures \ref{plot transposkink} and \ref{plot transition point thin shell} (plot on the left). In plot on the right of the figure \ref{plot transition point thin shell} we set $\omega=2$.

The limiting regime al late times is the BTZ black hole. By employing the second formula in (\ref{Lreg ads3 btz}), one finds that the equation (\ref{transition eq general}) for the transition curve in the configuration space given by $\ell_2$, $\ell_1$ and $\ell_0$ can be written as follows
\begin{equation}
\label{transition eq btz}
\frac{\sinh(\pi\ell_1/\beta_H)\,\sinh(\pi\ell_2/\beta_H)}{\sinh(\pi\ell_0/\beta_H)\, \sinh(\pi(\ell_1+\ell_2+\ell_0)/\beta_H)} 
\,=\, 1\;.
\end{equation}
Introducing $\omega$ through $\ell_2=\omega \ell_1$ as above and using the addition formulas for the hyperbolic functions, the equation (\ref{transition eq btz}) becomes
\begin{equation}
\label{btz transition eq}
\sinh^2\left(\frac{\pi \ell_0}{\beta_H}\right)
\left[\, B(\ell_1,\omega)
 +  C(\ell_1,\omega) \coth\left(\frac{\pi \ell_0}{\beta_H}\right) 
\right]
\,=\,1
\end{equation}
where we have defined
\begin{equation}
B(\ell_1,\omega) \equiv \coth\left(\frac{\pi \ell_1}{\beta_H}\right)  \coth\left(\frac{\pi \omega \ell_1}{\beta_H}\right) +1
\hspace{.8cm}
C(\ell_1,\omega) \equiv \coth\left(\frac{\pi \ell_1}{\beta_H}\right) +\coth\left(\frac{\pi \omega \ell_1}{\beta_H}\right) \;.
\end{equation}
Expressing the hyperbolic functions in their exponential form, the equation (\ref{btz transition eq}) becomes a second order equation in terms of $e^{2\pi\ell_0/\beta_H}$ and its positive root provides $\ell_0$ in terms of $\ell_1$ and $\omega$. The result reads 
\begin{equation}
\label{btz transition curve}
\ell_0 \,=\,\frac{\beta_H}{2\pi}\,\log\left(\frac{B(\ell_1,\omega) +2+\sqrt{4\big[1+B(\ell_1,\omega)\big]
 +  C(\ell_1,\omega)^2}}{B(\ell_1,\omega) +  C(\ell_1,\omega)}\, \right)\;.
\end{equation}
In the figures \ref{plot transposkink} and \ref{plot transition point thin shell} this curve is the black one below all the other ones.
The curve (\ref{btz transition curve}) passes through the origin $(\ell_1, \ell_0)=(0,0)$ and it always stays below the line (\ref{ads3 transition curve}). Moreover, the line (\ref{ads3 transition curve}) is tangent to (\ref{btz transition curve}) at the origin and this provides a check of (\ref{btz transition curve}) because for small $\ell_1$ and finite $\omega$ (which implies small $\ell_2$ as well) the minimal curves remain close to the boundary and therefore only the asymptotic geometry of BTZ, which is $AdS_3$, matters.
For any finite $\omega>0$, the curve (\ref{btz transition curve})  tends asymptotically to a horizontal line $\ell_0=\tilde{\ell}_0$ when $\ell_1$ is large for any finite value of $\omega$. 
In this limit both the disjoint interval are large while the ratio between them, being given by $\omega$, is kept fixed.
Quantitatively, since for $\ell_1 \rightarrow +\infty$ we have $B(\ell_1,\omega)  \rightarrow 2$ and $C(\ell_1,\omega)  \rightarrow 2$, the asymptotic value $\tilde{\ell}_0$ reads
\begin{equation}
\tilde{\ell}_0 \,=\, \frac{\beta_H}{\pi}\, \log \sqrt{2}
\end{equation}
and it is independent of $\omega$. This means that in the BTZ background, when the separation $\ell_0$ is larger than $\tilde{\ell}_0$, the holographic mutual information is zero for any $\ell_1$ and $\ell_2$.

\section {Strong subadditivity and  null energy condition}
\label{section SSA and NEC}

In this section we explore the relation between the null energy condition for the Vaidya metrics and the strong subadditivity condition, which is an important inequality satisfied by the entanglement entropy. We
find that a violation of the null energy condition leads to a violation of strong subadditivity\footnote{We remark that this result has been independently  obtained also by Robert Callan and Matthew Headrick.}.

\begin{figure}[th]
\begin{tabular}{c}
\vspace{.7cm}
\hspace{-0.4cm}
\includegraphics[width=3.in]{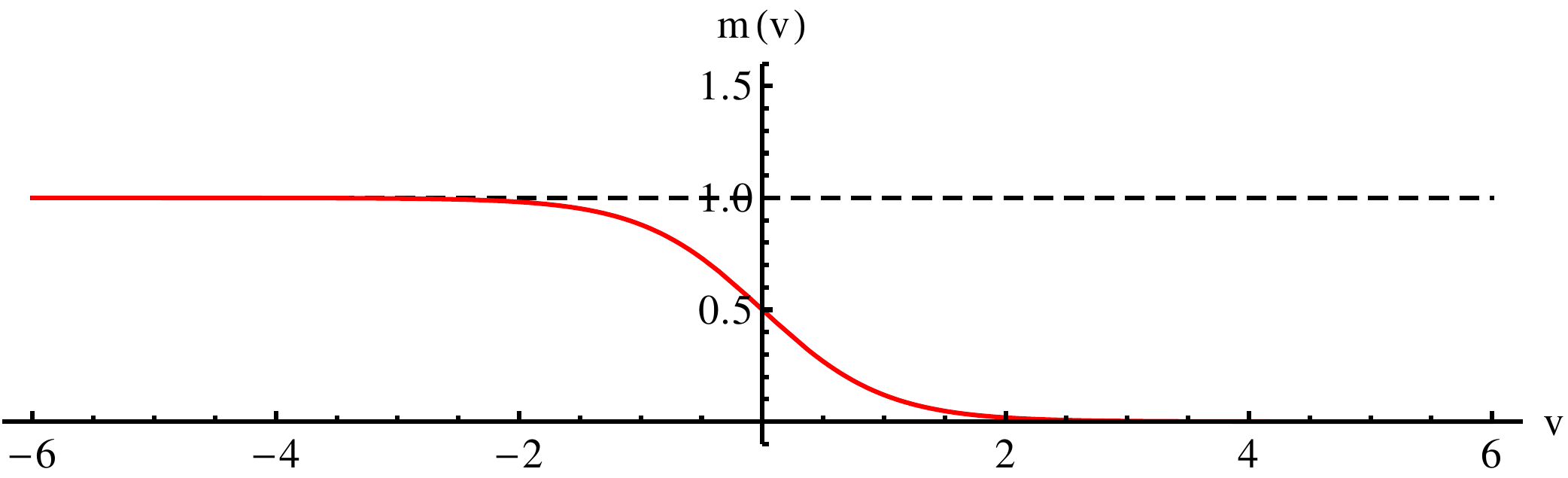}
 \hspace{+.6cm} 
\includegraphics[width=3.in]{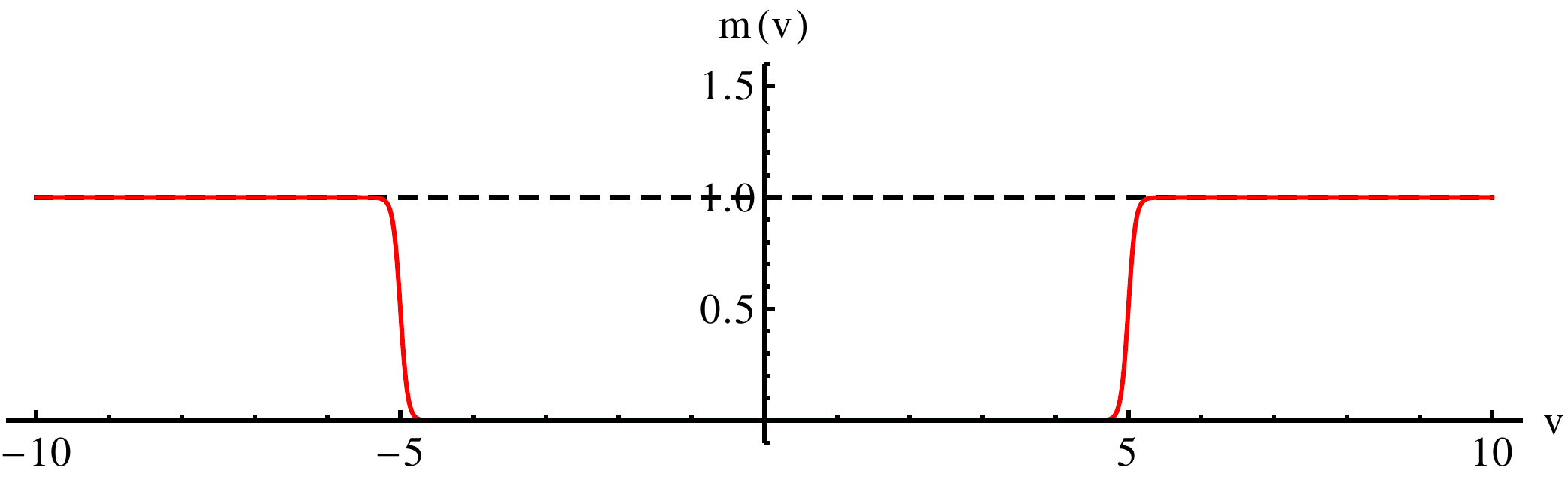}\\
\hspace{-0.8cm}\includegraphics[width=3.2in]{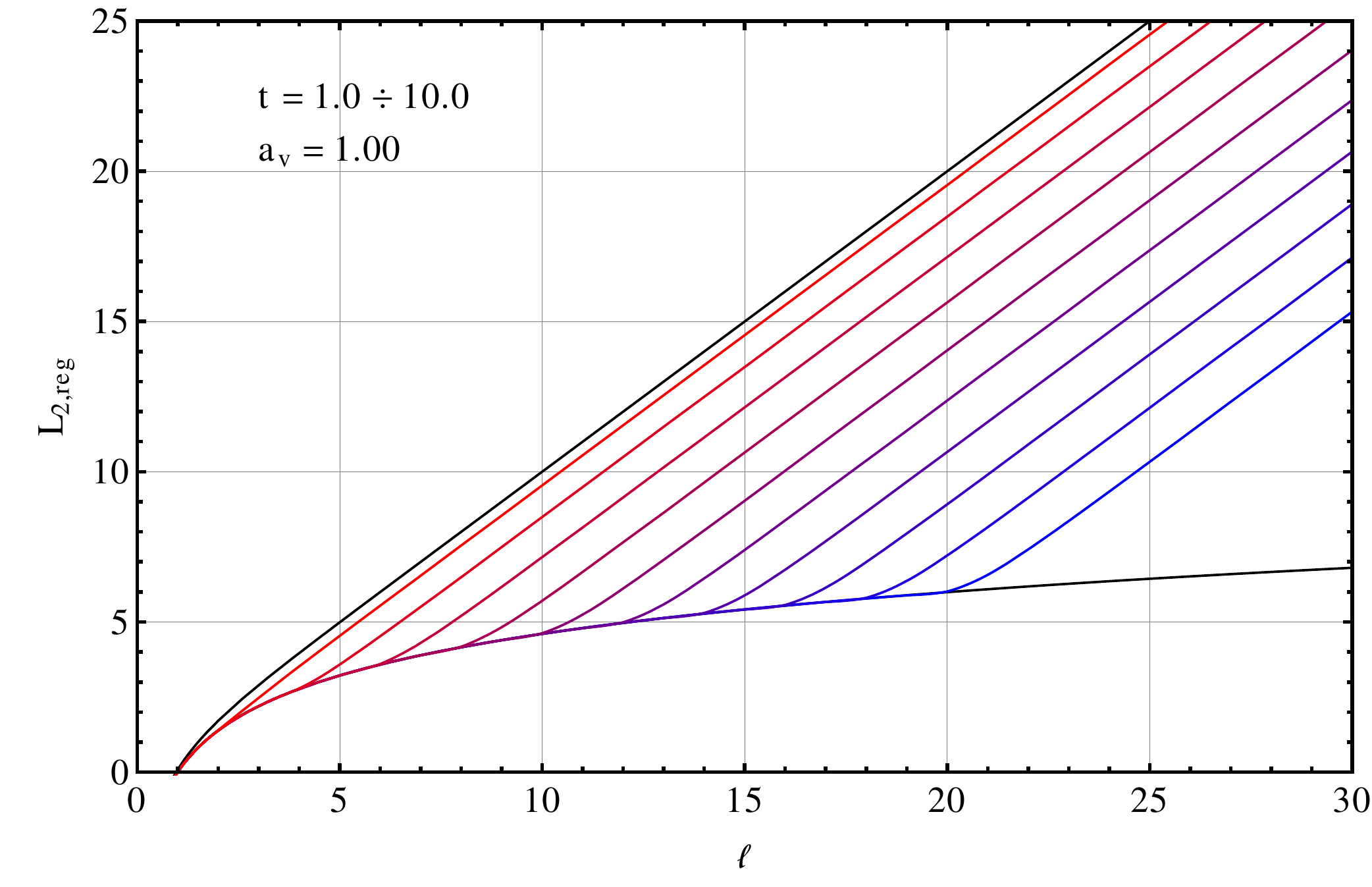}
 \hspace{-.0cm}  
\includegraphics[width=3.2in]{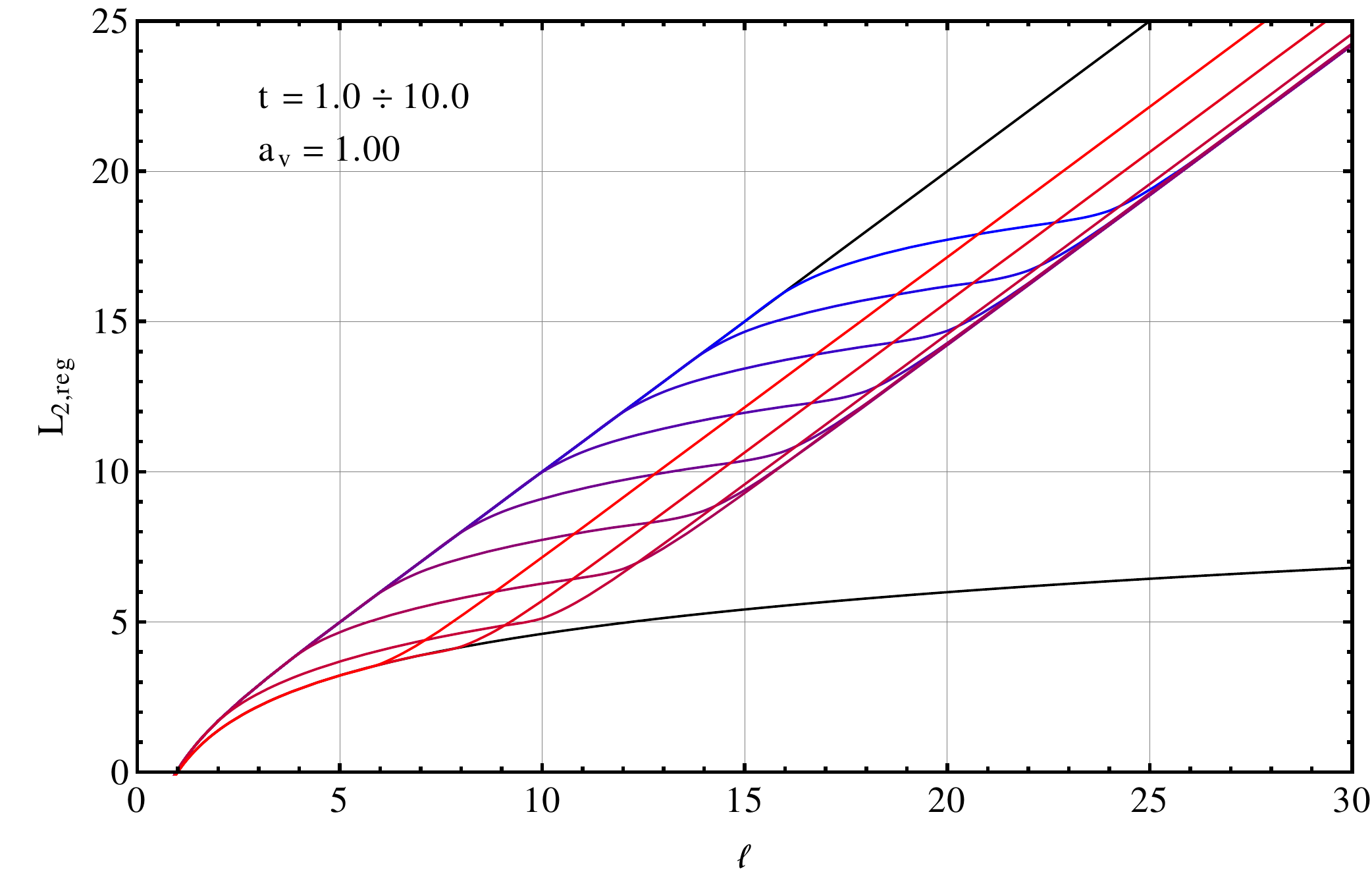}
\end{tabular}
\caption{Regularized length
$L_\mathrm{reg}$ for three dimensional Vaidya metric when the null energy condition is violated: a change of concavity is manifest in both cases. The mass profiles used are shown above and they are
 \mbox{$m(v) = \tfrac{M}{2} [2 + \tanh ((v - v_0/2)/a_v) - \tanh((v + v_0/2)/a_v)]$} on the right and \mbox{$m(v) = \tfrac{M}{2}[1-\tanh(v/a_v)]$} on the left, with $M = 1$, $a_v = 0.1$, $v_0 = 10$.
\label{plot Lreg nec violation}}
\end{figure}

Consider a quantum system that is partitioned into three or more subsystems, i.e. its Hilbert space  $H$  can be written as $ H = \otimes_i H_i$, and let us denote by $\rho_{i_1,i_2,\ldots}$ the reduced density matrix obtained by tracing the full density matrix of the system over all $H_j$ with $j \neq i_1, i_2,\ldots$. It can be shown, on very general grounds, that the Von Neumann entropy satisfies the {\it subadditivity condition} 
\begin{equation}
\label{subadditivity}
S(\rho_{1}) + S(\rho_{2}) \geqslant S(\rho_{1,2})
\end{equation} 
and also the following two inequalities
\begin{equation}
\label{strong subadditivity}
\begin{split}
 &S(\rho_{1,2}) + S(\rho_{2,3}) \geqslant S(\rho_2) + S(\rho_{1,2,3})\\
 &S(\rho_{1,2}) + S(\rho_{2,3}) \geqslant S(\rho_1) + S(\rho_{3})
\end{split}
\end{equation} 
which are equivalent and known as {\it strong subadditivity condition} (see \cite{NielsenChuang, Headrick:2007km, Hayden:2011ag} and the refs therein for more detailed discussions). 

If the Hilbert space is partitioned into the product of the Hilbert spaces of local degrees of freedom belonging to non intersecting regions of space $A_1, A_2,\ldots$, the inequalities (\ref{subadditivity}) and (\ref{strong subadditivity}) can be written respectively as
\begin{equation}
\label{SA}
S_{A_1} + S_{A_2} \,\geqslant \, S_{A_2 \cup A_2}  
\end{equation} 
and 
\begin{equation}
\label{SSA}
\begin{split}
&S_{A_1 \cup A_2} + S_{A_2 \cup A_3} \,\geqslant \, S_{A_2} + S_{A_1 \cup A_2 \cup A_3}\\
&S_{A_1 \cup A_2} + S_{A_2 \cup A_3} \,\geqslant \, S_{A_1} + S_{A_3}\;.
\end{split}
\end{equation}

\begin{figure}[th]
\begin{tabular}{c}
\hspace{-0.8cm}\includegraphics[width=3.2in]{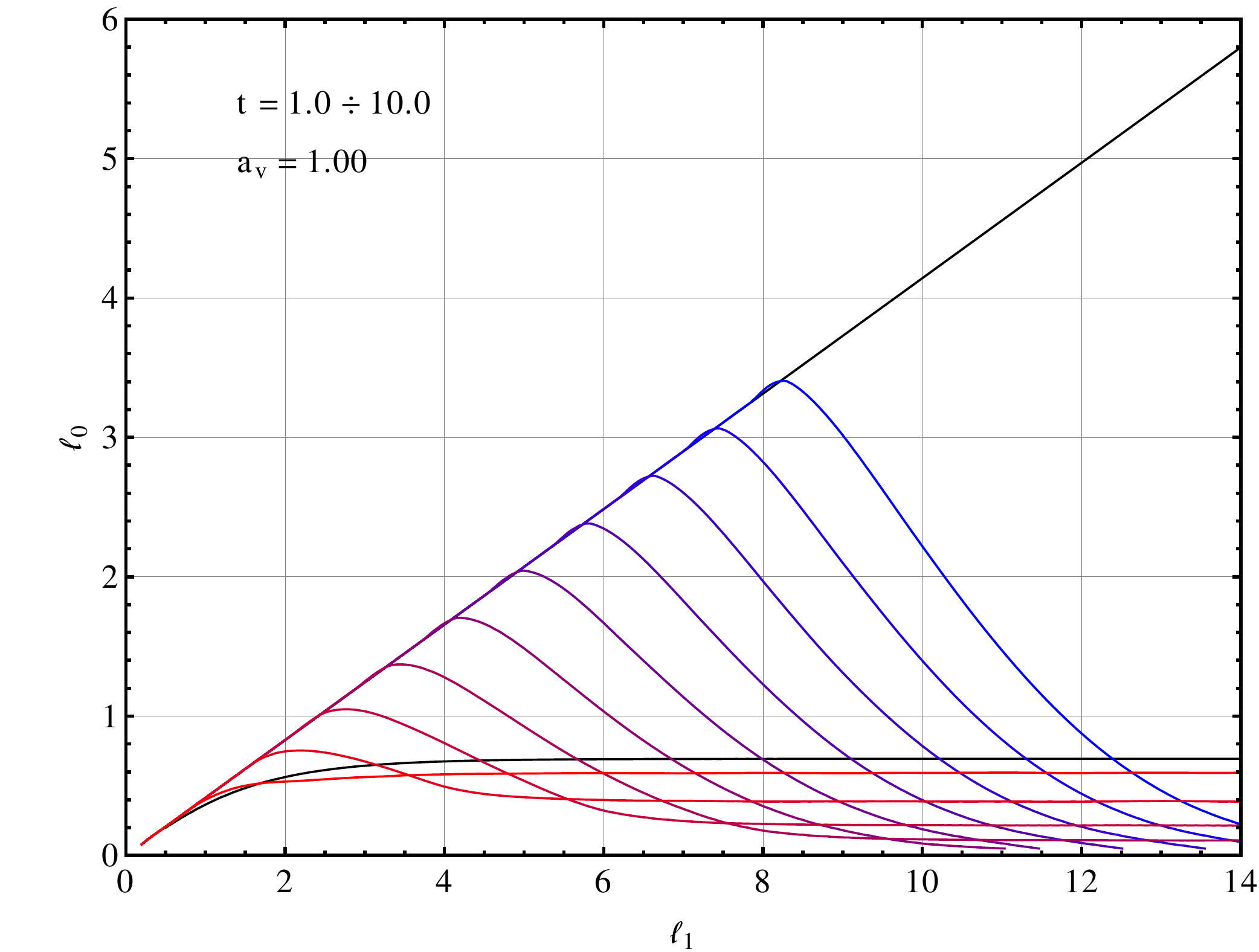}
 \hspace{-.0cm}  
\includegraphics[width=3.2in]{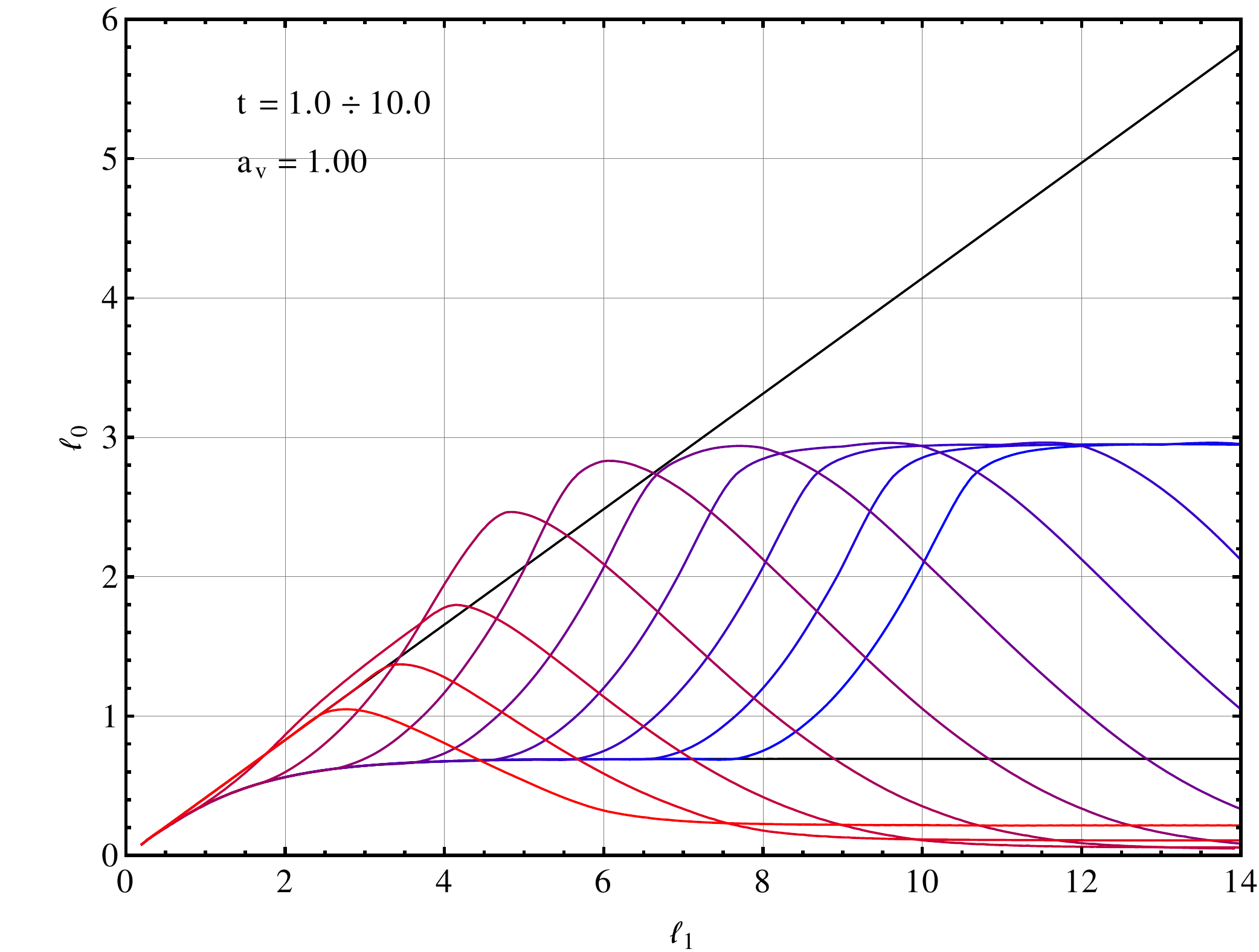}
\end{tabular}
\caption{
Location of the transition point of $I(\ell_0,\ell_1,\ell_1)$ when the null energy condition is violated: the curve is not monotonically increasing with $\ell_1$. The mass profiles employed in the plots are the same as in figure \ref{plot Lreg nec violation}.
\label{plot transition point nec violation}}
\end{figure}

\noindent In one dimensional systems (or when the symmetry of the regions considered is such that the problem is effectively one-dimensional), for a complete description, it is sufficient to consider the entanglement entropy  of an interval as a function of its length $\ell$, and the two inequalities of the strong subadditivity are more conveniently expressed in terms of the function $S(\ell)$.
The first inequality in (\ref{SSA}) states that the function $S(\ell)$ is concave, and the second that it is non decreasing.\\
In section \ref{subsection vaidya} we mentioned (see (\ref{pos mprime})) that for the Vaidya metrics (\ref{metric vadya d dim}) the condition $m'(v) \geqslant 0$ guarantees that the null energy condition is satisfied. 
By choosing a mass function that does not monotonically increase with $v$, we can violate the null energy condition and explore the consequences of this violation on the entanglement entropy. The results are shown in figure \ref{plot Lreg nec violation} and \ref{plot transition point nec violation}.  

\noindent
The curves in figure \ref{plot Lreg nec violation} are not concave functions of $\ell$, but they are still non-decreasing. Therefore only the first of the two inequalities is violated. This means that they cannot be equivalent in this setting. In order to clarify this apparent contradiction we have to discuss how the equivalence between the inequalities is proven, both in quantum mechanics and holographically.

\noindent 
The two inequalities can be shown to be equivalent by introducing an auxiliary fourth Hilbert space $H_4$ such that $\rho_{1,2,3} = \mathrm{Tr}_4 |\psi\rangle\langle\psi|$, for a certain pure state $|\psi\rangle$ \cite{NielsenChuang}. Then
\begin{equation}
 S(\rho_{1,2,4}) = S(\rho_3)
 \quad \text{and} \quad
 S(\rho_{1,4}) = S(\rho_{2,3})
\end{equation} 
and hence, if we write the first inequality for $3\leftrightarrow4$, $1\leftrightarrow2$
\begin{equation}\label{zero overall entropy}
  S(\rho_{1,2}) + S(\rho_{1,4}) \geqslant S(\rho_1) + S(\rho_{1,2,4})
\end{equation} 
and substitute, we get the second, and viceversa.\\
If one tries to replicate the argument above in the holographic setting, one encounters a difficulty  because, although we are guaranteed that it is always possible to find the Hilbert space $H_4$, it is not guaranteed that it will be the Hilbert space of the local degrees of freedom of some other region of the boundary theory. However, that is the only known kind of partitioning of the Hilbert space that allows for a holographic computation. It turns out that, if the bulk manifold is homologous to the boundary, the problem is easily solved:  $H_4$ can be taken to be the Hilbert space of the degrees of freedom of the region $A_4=\overline{A_1 \cup A_2 \cup A_3}$, the complement of $A_1 \cup A_2 \cup A_3$, because it satisfies
\begin{equation}\label{zero overall entropy holographic}
  S_{A_1 \cup A_2 \cup A_4} = S_{A_3}
\quad \text{and} \quad
 S_{A_1 \cup A_4 } = S_{A_2\cup A_3}
\end{equation} 
which are the equivalent of (\ref{zero overall entropy}).

On the other hand, if the manifold contains a black hole, the problem is more complicated, because the entropy of the entire system is non-zero, and hence (\ref{zero overall entropy holographic}) do not hold. 
One has to rely on the conjecture that the holographic entanglement entropy is actually describing the entanglement entropy of a certain quantum system, and hence on the quantum information proof mentioned above. If the dual geometry is sufficiently unphysical, as is the case for when the null energy condition is violated, it may not be the holographic description of any quantum system, and there is no a priori reason to expect that the two inequalities should be equivalent.

In terms of the mutual information $I(A_1, A_2)$ for two disjoint regions (\ref{MI def}) the subaddivity inequality (\ref{SA}) implies that 
\begin{equation}
  I(A_1, A_2) \geqslant 0
\end{equation} 
while the first strong subaddivity inequality (\ref{SSA}) can be written as follows
\begin{equation}
\label{SSA v2}
I(A_1, A_2 \cup A_3)\,\geqslant\,I(A_1, A_2)
\end{equation}
i.e., the mutual information increases as one of the two regions is enlarged while the other one is kept fixed. Applying this inequality twice, we one can also conclude that when two equal regions are enlarged by the the same quantity, the mutual information increases.

\noindent
From the transition curves in figure \ref{plot transition point nec violation}, we can observe that the holographic mutual information is not monotonically increasing with $\ell_1$. This behavior is another manifestation of the violation of the first strong subadditivity inequaliy, and has to be contrasted with figure \ref{plot transposkink} discussed in the previous section, where we used the the mass profile (\ref{mass pos kink}) (figure \ref{plot mass upkink}) and the null energy condition is satisfied.

\section{Holographic tripartite information and monogamy} 
\label{section I3}

In this section we consider the holographic tripartite information for three dimensional Vaidya metrics and show that the monogamy of the holographic mutual information is violated when the null energy condition is not satisfied. 

\begin{figure}[th]
\begin{center}
\includegraphics[width=5in]{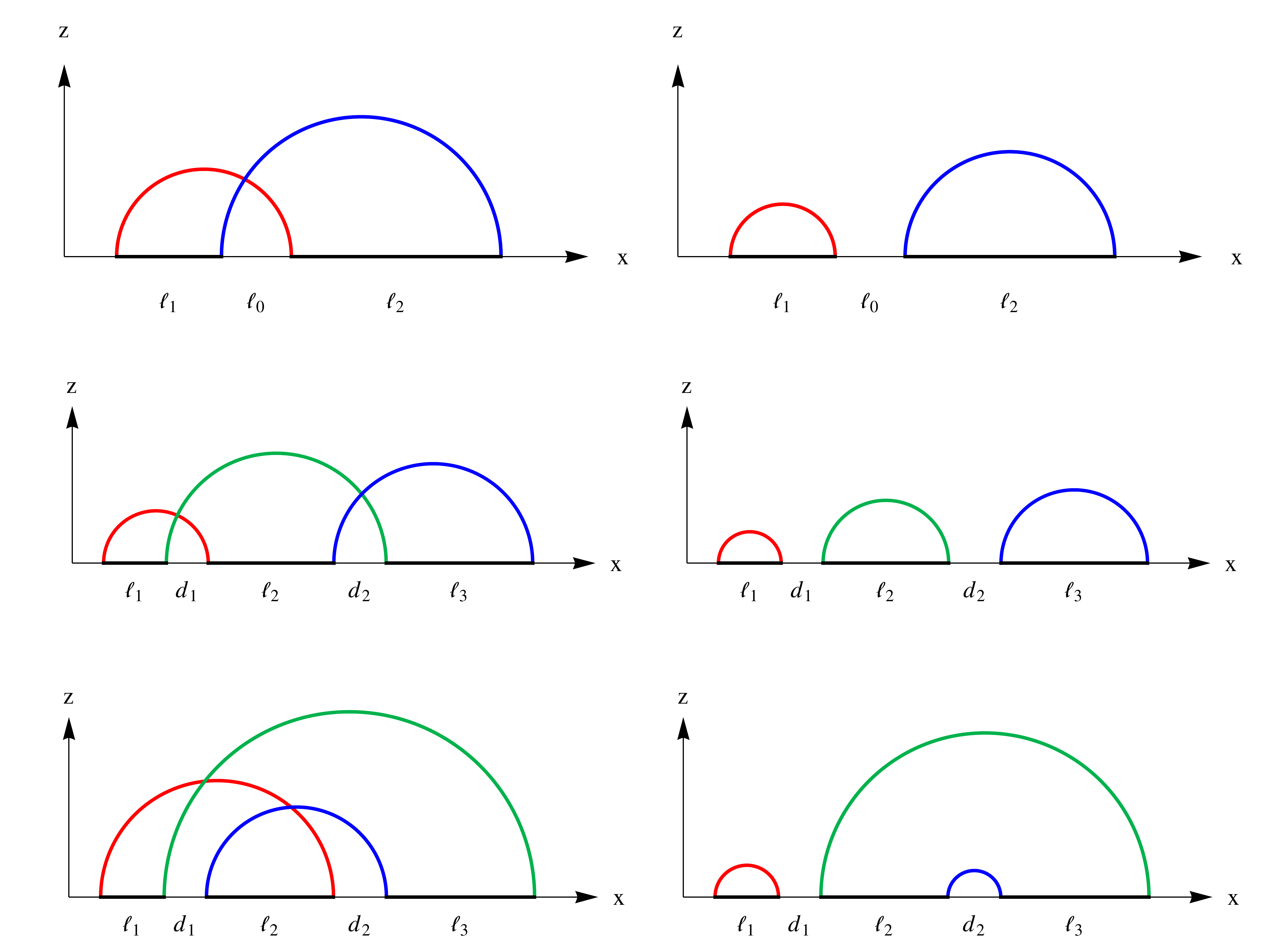} 
\end{center}
\vspace{-.6cm}
\caption{
\label{plot intersecting configs}
 Schematic representations of some mixed configurations occurring in the computation of $S_{A_1 \cup A_2}$ (upper line) and $S_{A_1 \cup A_2 \cup A_3}$ (middle and bottom line). If the regularized area of the surface homologous to a single region $A$ is an increasing function of the size of $A$, then the configuration on the left is suboptimal w.r.t. the one on the right (each line on the left should be compared with the one on the right having the same color) and thus it does not occur in the holographic mutual information or the holographic tripartite information.}
\end{figure}

In addition to the mutual information, another interesting quantity that can be defined from the entanglement entropy is the \emph{tripartite information}
\begin{equation}
\label{I3 def}
I_3(A_1, A_2, A_3) \,\equiv\,
S_{A_1} + S_{A_2}+S_{A_3}
- S_{A_1 \cup A_2}  - S_{A_1 \cup A_3}  - S_{A_2 \cup A_3}  
+  S_{A_1 \cup A_2 \cup A_3}
\end{equation}
where $A_1$, $A_2$ and $A_3$ are disjoint regions. In contrast with the mutual information, this quantity is free of divergences even when the regions share their boundary. It is a measure of the extensivity of the mutual information; indeed, the definition (\ref{I3 def}) can be written also as
\begin{equation}
 I_3(A_1, A_2, A_3) \,\equiv\,I(A_1,A_2) + I(A_1,A_3) - I(A_1,A_2\cup A_3)\,.
\end{equation} 
Thus, the mutual information is extensive when $I_3=0$, superextensive when $I_3<0$ and subextensive when $I_3>0$. In particular, in either the extensive or the superextensive case, namely
\begin{equation}
 I(A_1,A_2) + I(A_1,A_3) \leqslant  I(A_1,A_2\cup A_3)
\end{equation} 
the mutual information is said to be monogamous. For a generic quantum system, the tripartite information  can be positive, negative or zero, depending on the choice of the regions.\\
Recently it has been shown \cite{Hayden:2011ag} that for quantum systems with a holographic dual the tripartite information is always monogamous. As for the strong subadditivity condition, the proof only holds in the case of static dual geometries because in the dynamical backgrounds it is not always guaranteed that the surfaces involved in the mixed configurations intersect each other. It is therefore interesting to explore the behavior of the holographic tripartite information in the simple dynamical backgrounds like the three dimensional Vaidya geometries.

\begin{figure}[th]
\begin{center}
\includegraphics[width=6in]{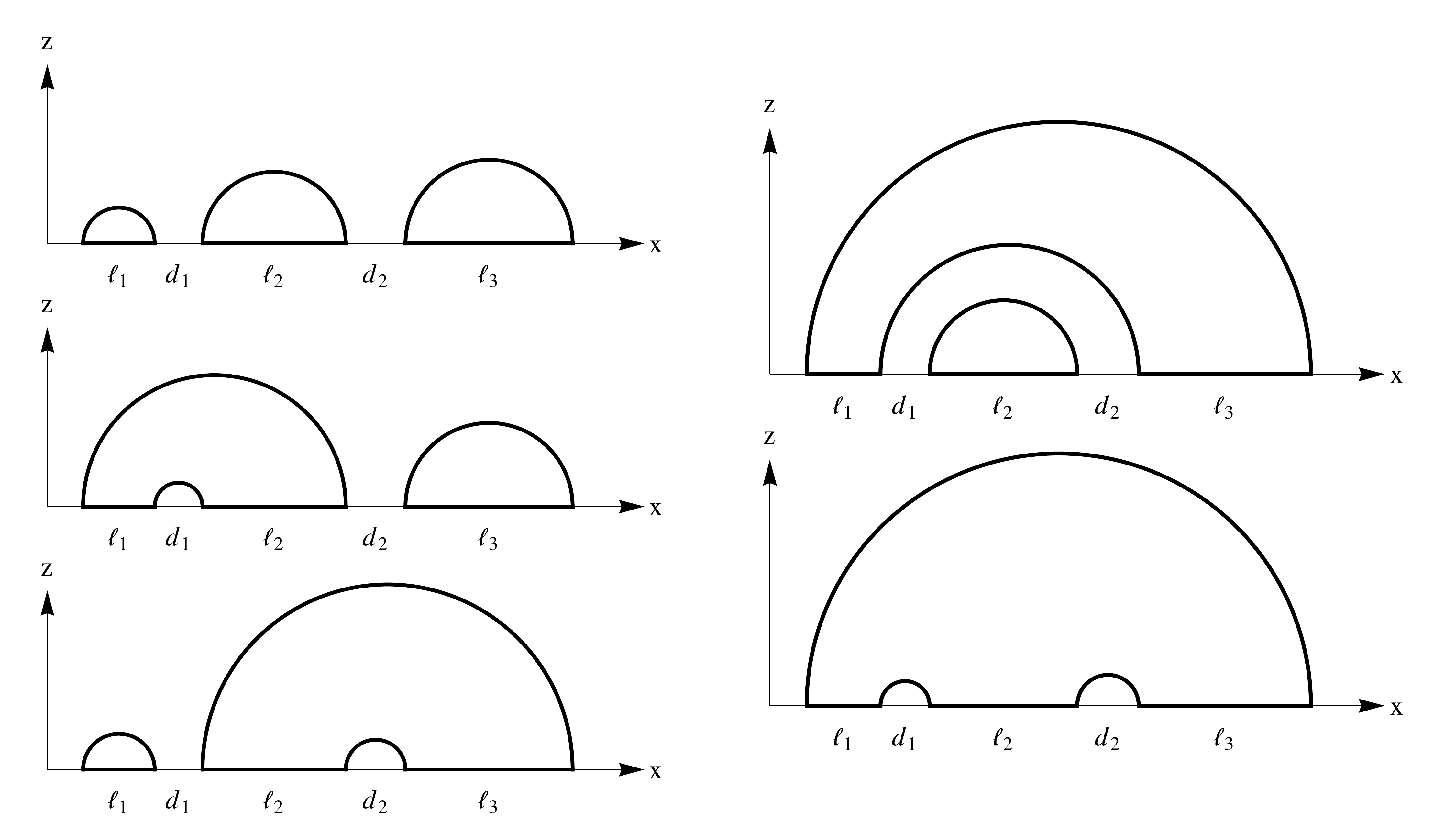} 
\end{center}
\vspace{-.6cm}
\caption{
\label{plot I3 geodesics configurations}
 Schematic geodesics configurations to consider in the computation of $S_{A_1 \cup A_2 \cup A_3}$.
}
\end{figure}

Among the terms occurring in the definition of the holographic tripartite information, the computation of  $S_{A_1 \cup A_2 \cup  A_3}$ deserves a short discussion. In one spatial dimension the three regions are just intervals, and, according to prescription of \cite{Hubeny:2007xt} for dynamical backgrounds, one has to find the extremal set of geodesics connecting all the extrema of the intervals. In principle, in presence of $N$ intervals one should compare $(2N-1)!!$ configurations (15 in our case). However, since $L_{2, \textrm{reg}}(\ell)$ at fixed time is an increasing function of $\ell$, for $N=3$ we are left only with the five configurations shown in figure \ref{plot I3 geodesics configurations}, by the argument sketched in the figure \ref{plot intersecting configs}. Thus, $S_{A_1 \cup A_2 \cup  A_3}$ is given by the minimum among the following quantities
\begin{equation}
\begin{array}{ll}
\begin{array}{l}
S(\ell_1) + S(\ell_2) + S(\ell_3) 
\end{array}
&  \textrm{three disconnected volumes}
\\
\left.
\begin{array}{l}
S(\ell_1+d_2+\ell_2) + S(d_1) + S(\ell_3) 
\\
S(\ell_1) + S(\ell_2+d_2+\ell_3) + S(d_2) 
\\
S(\ell_1+d_1+\ell_2+d_2+\ell_3) + S(d_1+\ell_2+d_2) + S(\ell_2) 
\end{array}\hspace{.4cm}\right\}
& \textrm{two disconnected volumes}
\\
\begin{array}{l}
S(\ell_1+d_1+\ell_2+d_2+\ell_3) + S(d_2) + S(d_2)  
\end{array}
& \textrm{one connected volume.}
\end{array}
 \end{equation}

\begin{figure}[th]
\begin{center}
\hspace{-.6cm}
\includegraphics[width=4.2in]{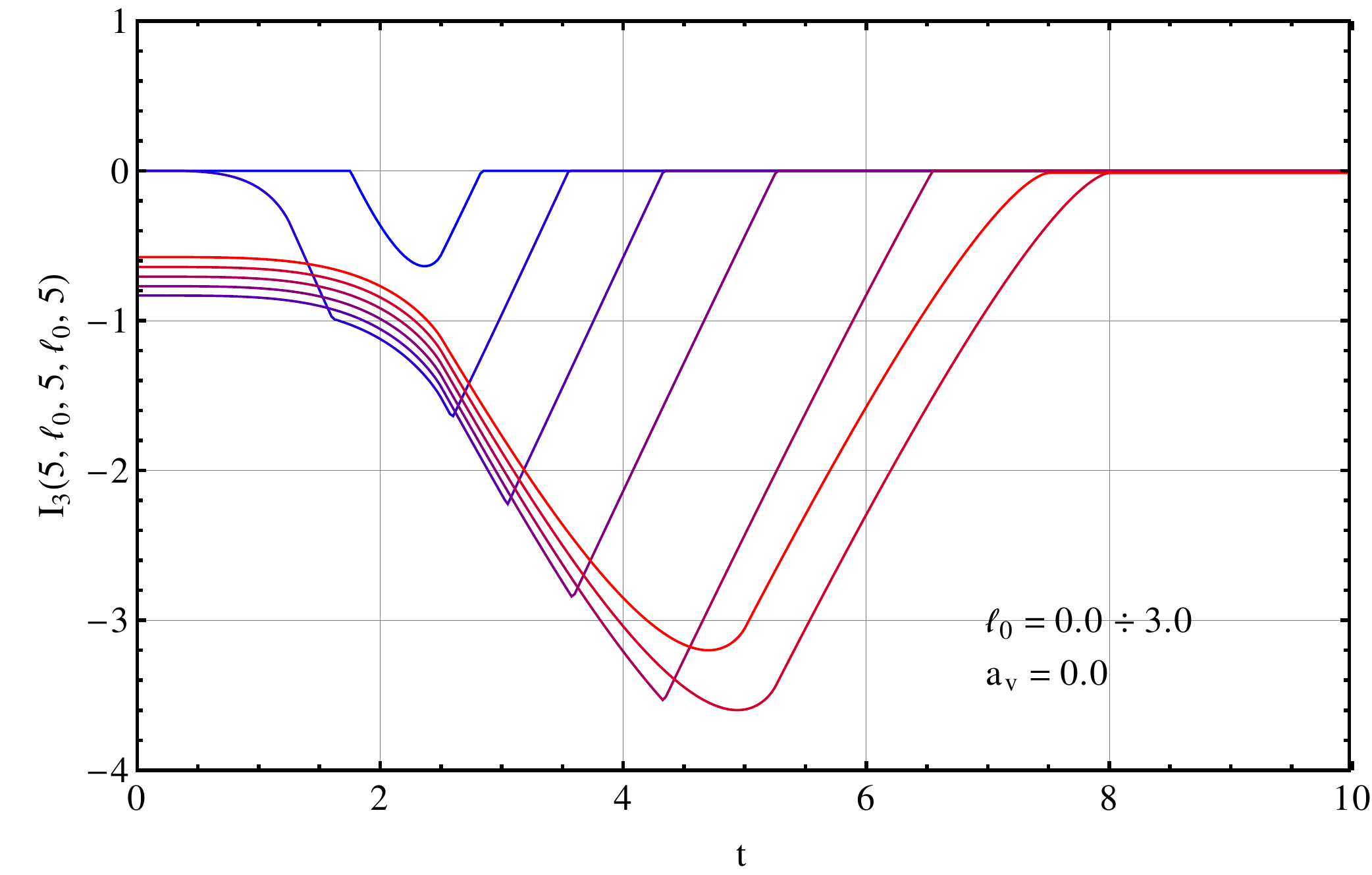}
\end{center}
\vspace{-.7cm}
\caption{Time evolution of the tripartite information for the three dimensional ($d=2$) Vaidya geometry in the thin shell limit. The three intervals have the same size $\ell_1 = \ell_2 = \ell_3 = 5$ and are separated by the same distance $d_1 = d_2 = \ell_0$. The plot shows that the holographic mutual information is always monogamous.
\label{plot I3 thin}}
\end{figure}

Figure \ref{plot I3 thin} displays the time dependence of the tripartite information when the intervals have the same size $\ell_1 = \ell_2 = \ell_3 = 5$ and are separated by the same amount $d_1 = d_2 = \ell_0$, with several values of $\ell_0$ shown. The behavior is quite complicated and involves different regimes, but the quantity is always non positive, even though the geometry is not static. At late times all curves go to zero, meaning that, in the thermal state dual to the BTZ black hole, the holographic mutual information is extensive.

\begin{figure}[th]
\begin{center}
\hspace{-.6cm}
\includegraphics[width=4.2in]{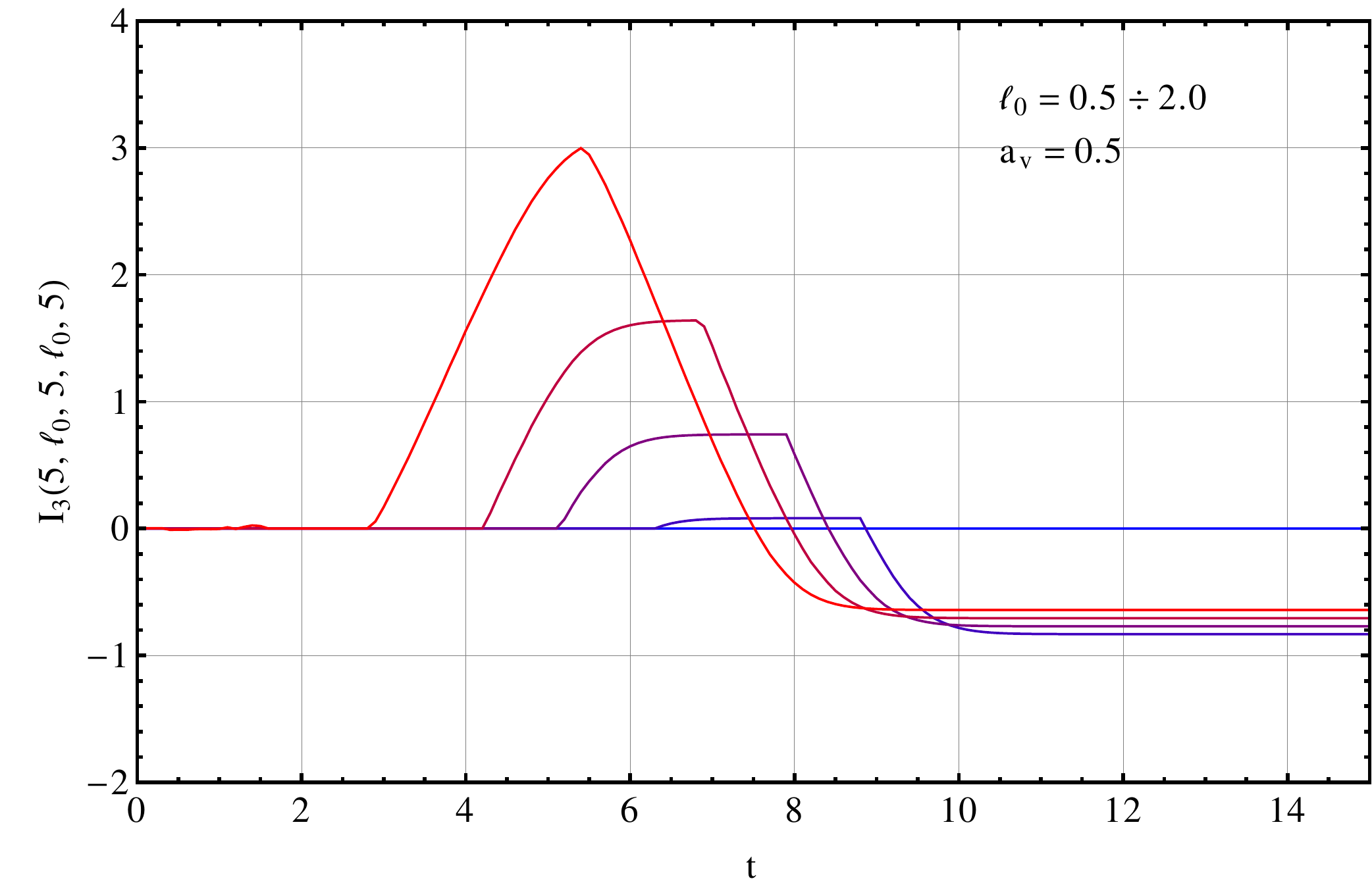}
\end{center}
\vspace{-.7cm}
\caption{Time evolution of the tripartite information for the three dimensional ($d=2$) Vaidya geometry when the null energy condition is violated: the mass function decreases from $m(-\infty) = 1$ to $m(+\infty) = 0$, according to the profile shown in the figure \ref{plot Lreg nec violation}. Since $I_3$ becomes positive for certain range of $t$, the monogamy of the holographic mutual information is violated.
\label{plot I3 violation NEC}}
\end{figure}

Given the results of the section \ref{section SSA and NEC} about the relation between the strong subadditivity of the holographic entanglement entropy and the null energy condition, we can explore the possible relation between the monogamy of the holographic mutual information and the null energy condition in the same way, namely by employing mass profiles $m(v)$ which have $m'(v) <0$ for some range of $v$. This is relevant because the strong subadditivity and the monogamy are indenpendent conditions. 
In the figure \ref{plot I3 violation NEC} we show the time evolution of the holographic tripartite information with the same interval configuration of the figure \ref{plot I3 thin} but with the mass function $m(v)$ decreasing from  $M = 1$ at early times ($v \rightarrow -\infty$) to $M= 0$ at late times ($v \rightarrow +\infty$) according to the profile shown in the figure \ref{plot Lreg nec violation} (plot on the left). The holographic tripartite information becomes positive for certain ranges of $t$, telling us that a violation of the null energy condition leads to a non monogamous holographic mutual information.

\section {Conclusions}

We studied the holographic mutual information for dynamical backgrounds given by the Vaidya metrics in three and four dimensions. 
We found that it is non monotonic as function of the boundary time and its behavior depends on the configuration of the two disjoint regions. 

\noindent 
From the transition curves of the holographic mutual information in the configuration space we could identify the different behaviors and also find a region in the configuration space where the holographic mutual information is zero at all times. Considering the holographic tripartite information, we observed that the holographic mutual information is monogamous also for these time dependent backgrounds in the ranges of the variables explored.

By modifying the mass profile occurring in the Vaidya metrics, we showed that the null energy condition is a necessary condition both for the strong subadditivity of the holographic entanglement entropy and for the monogamy of the holographic mutual information.
A deeper understanding of the relation between the null energy condition and the inequalities satisfied by the quantities defined from the holographic entanglement entropy is needed.

\subsection*{Acknowledgments}

We acknowledge Francesco Bigazzi, John Cardy, Pasquale Calabrese, Horatio Casini, Mark Mueller, Andrea Sportiello and Tadashi Takayanagi for stimulating discussions.
We are grateful in particular to Matthew Headrick and John McGreevy for helpful conversations and comments.\\
ET acknowledges the organizers of the Aspen workshop {\it Quantum Information in Quantum Gravity and Condensed Matter Physics} and also Pisa University and the Institute Henri Poincar\'e for the warm hospitality during parts of this work.
ET is supported by Istituto Nazionale di Fisica Nucleare (INFN) through the ``Bruno Rossi'' fellowship.
AA and ET are supported by funds of the U.S. Department of Energy under the cooperative research agreement \mbox{DE-FG02-05ER41360}.


\end{document}